\documentclass[12pt,preprint]{aastex}

\slugcomment{To appear in ApJSuppl (2003), 145, 259.  Full catalog figures
available in published article or at 
http://astro.berkeley.edu/\~\ thelfer/bimasong\_supplement.pdf .}
\shorttitle{BIMA Survey of Nearby Galaxies}
\shortauthors{Helfer et al.}

\begin{document}
\title{The BIMA Survey of Nearby Galaxies (BIMA SONG). \\
II. The CO Data
}

\author{Tamara T. Helfer\altaffilmark{1},
Michele D. Thornley\altaffilmark{2},
Michael W. Regan\altaffilmark{3},
Tony Wong\altaffilmark{1,4}, \\
Kartik Sheth\altaffilmark{5,6}, 
Stuart N. Vogel\altaffilmark{5},
Leo Blitz\altaffilmark{1},
and
Douglas C.-J. Bock\altaffilmark{1}
}
\altaffiltext{1}{Radio Astronomy Laboratory, 601 Campbell Hall, University of 
California, Berkeley, CA 94720; thelfer@astro.berkeley.edu}
\altaffiltext{2}{Department of Physics, Bucknell University, Lewisburg, PA 17837}
\altaffiltext{3}{Space Telescope Science Institute, 3700 San Martin Drive,
Baltimore, MD 21218}
\altaffiltext{4}{Current address: CSIRO Australia Telescope National Facility, 
P.O. Box 76, Epping, NSW 1710 Australia}
\altaffiltext{5}{Department of Astronomy, University of Maryland, College 
Park, MD 20742}
\altaffiltext{6}{Current address: California Institute of Technology, MS 105-24, Pasadena, CA
91125}

\date{\centering\today}

\begin{abstract}

The BIMA Survey of Nearby Galaxies is a systematic imaging study of the 
3 mm CO J = 1--0 molecular
emission within the centers and disks of 44 nearby spiral galaxies.
The typical spatial resolution of the survey is 6\arcsec\, or 360 pc at the
average distance (12 Mpc) of the sample.
The velocity resolution of the CO observations is 4 km~s$^{-1}$,
though most maps are smoothed to 10 km~s$^{-1}$ resolution.
For 33 galaxies, multi-field observations
ensured that a region $\ga$ 190\arcsec\ ($<$D$>$ = 10 kpc) in diameter was 
imaged.  For the remaining 11 galaxies, which had smaller optical
diameters and were on average farther away, 
single-pointing observations imaged a
100\arcsec-diameter ($<$D$>$ = 11 kpc) region.
The sample was not chosen based on CO or infrared
brightness; instead, all spirals were included that met the selection
criteria of $v_{\sun} \le$ 2000 km~s$^{-1}$, $\delta \ge$ -20\arcdeg,
$i \le$ 70\arcdeg, $D_{25} <$ 70\arcmin, and $B_T <$ 11.0.
The detection rate was 41/44 sources or 93\%; of the three nondetections,
one (M 81) is known to have CO emission at locations outside 
the survey field of view.
Fully-sampled single-dish CO data were
incorporated into the maps for 24 galaxies; these single-dish data
comprise the most extensive collection of fully-sampled,
two-dimensional single-dish CO maps of external galaxies to date.
We also tabulate direct measurements of the global CO flux densities
for these 24 sources.
For the remaining 20 sources, 
we collected sensitive single-dish spectra in order to
evaluate the large-scale flux recovery.
We demonstrate that the measured ratios of flux density recovered
are a function of the signal-to-noise of the interferometric data.
We examine the degree of central peakedness of the molecular surface
density distributions and show that the distributions exhibit
their brightest CO emission within the central 6\arcsec\ in
only 20/44 or 45\% of the sample.
We show that all three Local Group spiral galaxies have CO
morphologies that are represented in SONG, though the
Milky Way CO luminosity is somewhat below the SONG average,
and M31 and M33 are well below average.
This survey provides a unique public database 
of integrated intensity maps, channel maps, spectra, and 
velocity fields of molecular emission in nearby galaxies.
It also lays the groundwork for extragalactic
surveys by more powerful future millimeter-wavelength interferometers like 
CARMA and ALMA.

\end{abstract}


\section{Introduction}

In the twenty-five years since the first detection of carbon monoxide in
external galaxies \citep{Rickard75}, astronomers have observed the 3 mm
J = 1--0 transition of CO in the centers of several hundreds of galaxies
\citep{Verter90, YS91, Young95, Casoli98, Nishiyama01, Paglione01}; 
these observations form
the basis for our statistical understanding of the molecular content
of galaxies as a function of parameters like galaxy type and luminosity.
Despite an abundance of studies of the amount of CO in galaxies, however,
there is still a rather incomplete understanding of the distribution
of CO within galaxies.   This is because the overwhelming majority of 
CO observations have been made using single-dish 
telescopes, the linear resolutions of which are often ten to one hundred 
times larger than the basic unit in which molecular gas is thought to be
organized, the giant molecular cloud (GMC).   Also, even for the roughly 200
galaxies that have been observed in multiple pointings with single-dish
telescopes, all but a handful have been observed only at select positions 
along the major and minor axes, rather than in fully-sampled, 
two-dimensional maps.  So far, only a few dozen galaxies have been
imaged with the high resolution achievable with an interferometer.  
The largest published interferometric
survey to date is by \citet{Sakamoto99}, who surveyed the central arcminute
of 20 nearby spiral galaxies with 4\arcsec\ resolution using telescopes
at the Nobeyama Radio Observatory (NRO) and the Owens Valley Radio
Observatory (OVRO).

The BIMA Survey of Nearby Galaxies (BIMA SONG) is the first systematic
imaging survey of the 3 mm CO J = 1--0 emission from the centers and
disks of nearby spiral galaxies.  The basic database that we have produced
is a collection of spatial-velocity data cubes for 44 nearby spiral 
galaxies that show the 
distribution and kinematics of CO emission at resolutions of a few hundred 
parsecs ($\sim$ 6\arcsec) and $\sim$ 10 km~s$^{-1}$ over a field of view of
typically 10 kpc ($\sim$ 190\arcsec).
BIMA SONG was introduced in the
recent study by \citet{Regan2001}, hereafter Paper I, where
we presented initial results for a subsample of 15 spiral galaxies.
In this paper, we present data from the full survey of 44 spiral
galaxies.  Some of the notable features of the BIMA SONG survey are as
follows:
(1) The sources were not selected based on 
CO brightness or infrared luminosity.
(2) We mosaiced multiple fields for the 33 galaxies with the largest 
optical diameters, so that the full-width half-power field of view 
was 190\arcsec.  This field of view is a full order of magnitude larger 
in area than was obtained for previous high-resolution surveys.
For the 11 galaxies with smaller optical diameters, 
we used a single pointing of the
interferometer to image a region 100\arcsec\ in diameter.
(3) The data have very good sensitivities, with rms noise levels of
23--130 mJy~bm$^{-1}$ per 10 km~s$^{-1}$ channel.  These are comparable
to the sensitivities obtained by \citet{Sakamoto99} of 17--110 mJy~bm$^{-1}$ 
per 10 km~s$^{-1}$ channel over a much smaller field of view.  The noise
levels are roughly an order of magnitude lower than those of the most
extensive single-dish survey, the FCRAO Extragalactic CO Survey
\citep{Young95}.
(4) We collected fully-sampled, On-the-Fly (OTF) 
single-dish maps for 24 of the brightest
galaxies and incorporated these data into the SONG maps, so that all of the CO
flux is imaged for these galaxies.  These maps in
themselves comprise the most extensive collection of fully-sampled,
two-dimensional single-dish CO maps of external galaxies to date.  For the
remaining 20 sources, we collected sensitive single-dish spectra 
in order to evaluate the large-scale flux recovery.
(5) All observations, data reduction and analysis were carried out
using uniform procedures, so that
systematic differences among the resultant maps should be minimized.
(6) We are collecting complementary broad-band optical, infrared,
H$\alpha$, HI, and radio continuum data so as to have a parallel
set of observations for each galaxy.  Where available, we have
compiled existing data; otherwise, we are obtaining our own new
observations.  These complementary data will be presented in future
SONG papers.

This paper is organized as follows.  In \S 2, we list the sample
selection criteria and describe the basic properties of the sample.  
In \S 3, we give details of the interferometric
and single-dish observations.  In \S 4, we describe the data reduction
and calibration.  
In \S 5, we present a catalog of the BIMA SONG
data in various forms, including channel maps, integrated intensity maps, 
spectra, and basic velocity fields.  
In \S 6, we discuss the CO detection rate for the survey.
In \S 7, we present direct measurements of global fluxes for the
24 OTF sources in the survey.
In \S 8, we discuss the issue of large-scale flux recovery and present
measurements for the BIMA SONG sources.
In \S 9, we examine the degree of central peakedness of the molecular surface
density distributions.
In \S 10, we consider how the three Local Group spiral galaxies compare
with the nearby spirals presented here.
We summarize the conclusions in \S 11.  
Finally, we have included two appendices in which we give brief 
descriptions of some deconvolution issues at millimeter wavelengths
as well as techniques for combining single-dish data 
with interferometric data.  We note that the
data collection, reduction, and calibration procedures were described in 
considerable detail in Paper I, which is very much a companion to 
the present paper.  Although we present some of the same material here,
we have tried to emphasize somewhat different points in the two papers, 
and we encourage the reader to look to both references for a complete 
discussion of technical details of the survey.
The data from this survey are publicly available (see \S 11).

\section{ The BIMA SONG Sample}

\subsection{ Selection Criteria}

Sources were selected without reference to
CO or infrared brightness.  Instead, Hubble type Sa through Sd galaxies 
were selected with $v_{\sun} <$ 2000 km~s$^{-1}$; if $H_o$=75 
km~s$^{-1}$~Mpc$^{-1}$, this corresponds to galaxies with Hubble distances
of $d \la$ 27 Mpc.  The galaxy inclinations, taken as cos $i$ = $R_{25}^{-1}$
from axial ratios $R_{25}$ in the RC3 \citep{RC3}, were $i \le$ 70\arcdeg, so
as not to be too edge-on to study azimuthal properties.
A minimum declination of $\delta \ge$ -20\arcdeg\ ensured that the
sources could be observed well from the Hat Creek Radio Observatory in
northern California without having too elongated a synthesized beam.
To limit the size of the sample,
we chose galaxies with apparent total blue magnitudes $B_T <$ 11.0.
Finally, we used a maximum optical diameter $D_{25} <$ 70\arcmin\ 
to exclude one source, M33, which is the subject of a separate BIMA
survey \citep{Engargiola02}.   The remaining sources have optical
diameters $D_{25} <$ 30\arcmin.
Using the NASA/IPAC Extragalactic Database (NED)\footnote{NED is operated
by the Jet Propulsion Laboratory, Caltech, under contract with the
National Aeronautics and Space Administration.}, we identified 44
galaxies meeting all selection criteria.  Table 1 lists these 44
sources, which comprise the BIMA SONG sample.

\subsection{ Properties of the Sample }

\begin{figure}
\centerline{\includegraphics[height=4in]{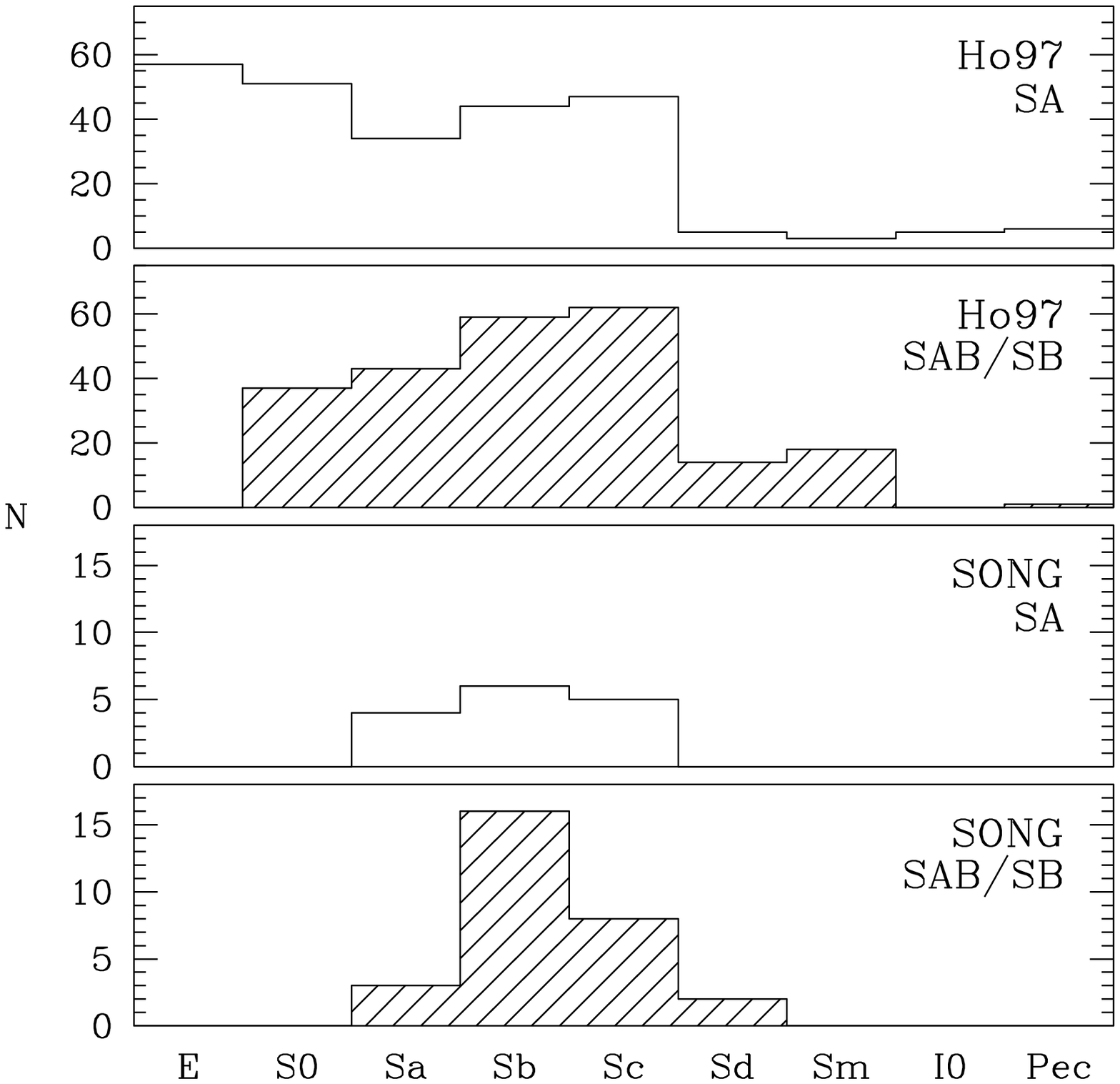} }
\caption{Distribution of RC3 Hubble types compared with RC3 types from
the Palomar survey
\citep{HFS97}, a magnitude-limited survey of the northern sky.  
The SONG selection was restricted to spirals of type
Sa -- Sd, while the 486 galaxies in the \citet{HFS97} sample
included earlier and later types.
The SONG sample has relatively few Sa/Sab galaxies; this is
probably a selection effect, since early type spirals tend to be 
found in clusters and are thus farther away on average than the 
SONG selection criteria allowed.  
}
\label{typehist}
\end{figure}

\begin{figure}
\centerline{\includegraphics[height=4in]{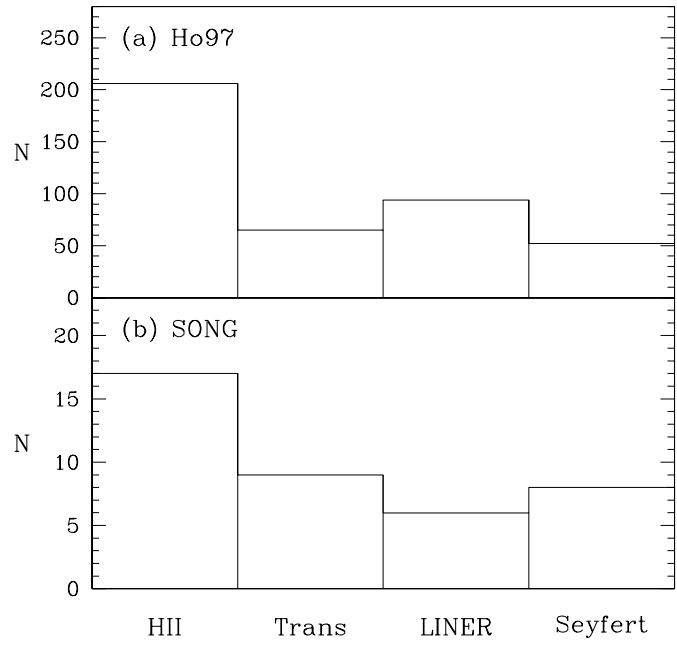} }
\caption{ Distribution of SONG nuclear classifications compared with 
those from the Palomar survey
\citep{HFS97}.   The similarity in the distributions  
reflects the similarities in the selection criteria for the two
samples. 
}
\label{nuchist}
\end{figure}

\begin{figure}
\centerline{\includegraphics[height=4in]{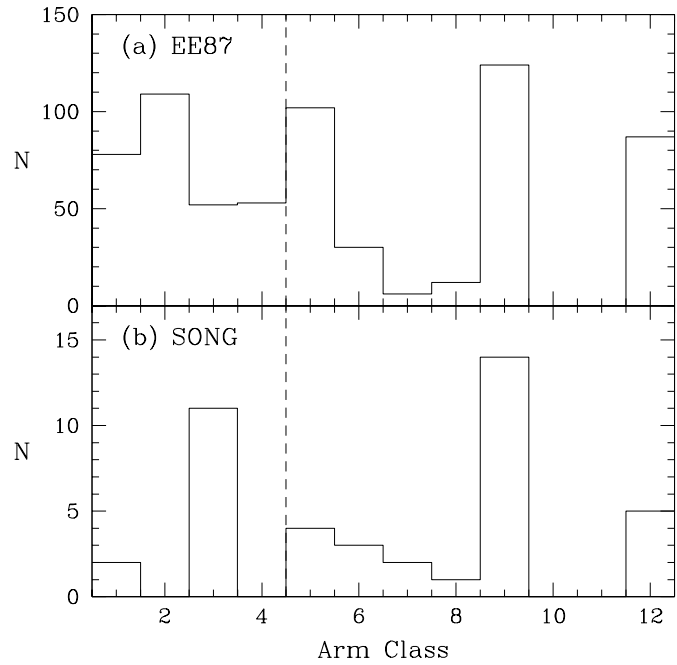} }
\caption{Distribution of arm classes (AC) compared with ACs from 654 spiral
galaxies in \citet{EE87}.
AC 1 -- 4 (left of the vertical dashed line) are ``flocculent''
galaxies, whereas AC 5 -- 12 are ``grand design'' spirals.
The SONG sample has relatively few galaxies in AC 1 -- 5 compared with 
the \citet{EE87} sample; this may be because the latter sample has
more low-luminosity galaxies than does SONG.
}
\label{armhist}
\end{figure}

The BIMA SONG sample includes objects with a range of properties
that are of interest for a variety of studies.
Spirals with Hubble classifications from Sab to Sd are represented,
with roughly one-third of the sample having Hubble type of Sbc and
the remaining objects split evenly between earlier and later types
(ab:b:bc:c:cd:d = 7:8:14:7:6:2). In Figure \ref{typehist},
we show the distribution of RC3 Hubble types compared with
the 486 galaxies from the magnitude-limited Palomar survey of the
northern sky \citep{HFS97}.  Compared to the Palomar survey,
the SONG sample has relatively few Sa/Sab galaxies.  This is
probably a selection effect, since early-type galaxies tend to
be found in clusters and are thus farther away on average than
the SONG selection criteria allowed.  The median distance for
the \citet{HFS97} sample is 17.9 Mpc, compared with 11.9 Mpc for
SONG.
Nearly two-thirds of the sample are
classified optically in the RC3 as barred galaxies, with five SB and 
24 SAB bars; the remaining 15 galaxies are optically unbarred.  
Using the nuclear classifications from the \citet{HFS97} survey,
we note that fourteen of the
sources contain nonstellar (Seyfert or LINER) emission-line nuclei,
seventeen have starburst (H II) nuclei, and nine show ``transition'' spectra,
with [O I] emission-line strengths intermediate between those of
H II nuclei and LINERs.  As Figure \ref{nuchist} shows, the
nuclear types are distributed in a similar way to those from the
Palomar survey.  
In Figure \ref{armhist}, we show the distribution of arm classifications
(AC) as defined (and classified) by \citet{EE87}, 
compared with the \citet{EE87} full sample 
of 654 spiral galaxies.  There are 29 grand-design spirals (AC = 5--12),
of which 19 show long arms dominating the optical disk (AC = 9, 12);
13 spirals are flocculents (AC = 1--4), exhibiting fragmented arms
that typically lack regular spiral structure.  The SONG sample
has relatively few galaxies in AC 1 -- 5 compared with the
\citet{EE87} sample; this may be because the latter sample
has more low-luminosity galaxies than does SONG.
The galaxies range in distance from 2.1 Mpc to nearly 26 Mpc, with
a mean distance of 12.2 $\pm$ 5.9 Mpc.  
The properties of the galaxies are summarized in Table 1.

\section{ Observations }

\subsection{ BIMA Observations}

We carried out BIMA SONG observations from 1997 November through 1999 
December using the 10-element Berkeley-Illinois-Maryland Association (BIMA)
millimeter interferometer \citep{Welch96} at Hat Creek, CA.  
In all, we included data from about 160 8-hour BIMA SONG tracks as well
as about 15 tracks on SONG sources taken by individual team members 
before 1997 November 
\citep{Regan99,Wong2000,WongBlitz2000}.
Before 1998 September, the antennas were configured in the then-named C array,
which included baselines as short as 7.7 m and as long as 82 m.
After 1998 September, we used the current C and D configurations, which
together include baselines as short as 8.1 m and as long as 87 m.
For all galaxies, we observed the CO J = 1--0 line at 115.2712 GHz.
We configured the correlator to have a resolution of 1.56 MHz (4 km~s$^{-1}$)
over a total bandwidth of 368 MHz (960 km~s$^{-1}$).

We observed 33/44 of the galaxies (those with R$_{25}$ $>$ 200\arcsec\
as well as three smaller sources) 
using a 7-field, hexagonal mosaic with a spacing of $\lambda/2D$ or
44\arcsec; this yielded a half-power field of view of about 190\arcsec\
or 10 kpc at the average distance of galaxies in the survey.
Of these, we observed some galaxies (NGC~0628, NGC~1068, NGC~2903, NGC~3627, 
NGC~4736, NGC~5033, NGC~5194) with slightly different pointing 
spacings or additional fields.
The remaining 11/44 galaxies were observed with a single pointing,
which yielded a field of view of 100\arcsec\ FWHM.  
For the multiple-pointing observations, each field was observed for one 
minute before the telescope slewed to the next pointing.  Thus, we returned 
to each pointing after $\sim$8 minutes, which is within the time needed 
to ensure Nyquist radial sampling of even the longest baselines in 
the $uv$ plane; see \citet{Helfer2002}.

Table 2 summarizes the observational parameters of the maps, including the
number of fields observed, the synthesized beam size in angular and
linear dimensions, the velocity binning, and the measured rms noise level 
in the channel maps.

\subsection{ NRAO 12 m Observations}

We collected over 700 hours of single-dish data from
1998 April through 2000 June using the NRAO 12 m telescope on Kitt
Peak, AZ\footnote{The National Radio Astronomy Observatory is operated
by the Associated Universities, Inc., under cooperative agreement with the
National Science Foundation.}.   We observed orthogonal polarizations
using two 256 channel filterbanks at a spectral resolution
of 2.0 MHz (5.2 km~s$^{-1}$), and with a 600 MHz configuration of the digital 
millimeter autocorrelator with 0.80 MHz  (2.1 km~s$^{-1}$) resolution as
a redundant backend on each polarization.  The pointing was monitored every 
1-2 hours with observations of planets and strong quasars.  The focus was 
measured at the beginning of each session and after periods during which
the dish was heating or cooling. 
At 115 GHz, the telescope half-power beamwidth, main beam efficiency
$\eta_m$, and forward spillover and scattering efficiency
$\eta_{fss}$ are 55\arcsec, 0.62, and 0.72.  To convert from
the recorded $T_R^*$ to $T_{MB}$, the reader should multiply
$T_R^*$ by $\eta_{fss}$/$\eta_m$ = 1.16.  To convert from
brightness temperature $T_R^*$ to units of flux density, the reader 
should multiply by the antenna gain, or 33 Jy K$^{-1}$.

For 24 galaxies (Table 2) with the strongest CO brightness temperatures, we
observed in On-the-Fly (OTF) mode \citep{Em96}, where the telescope takes 
data continuously as it slews across the source.   
The resultant OTF maps were 
incorporated into the BIMA maps to create combined BIMA+12m maps
that did not have large-scale emission 
``resolved out'' by the interferometer-only observations.  
The sources with combined maps are identified in the last column of Table 2.
For the remaining 20 galaxies, we obtained sensitive
position-switched spectra in order to evaluate the large-scale flux
recovery.  We describe the two observing modes in the following subsections.

\subsubsection{ On-the-Fly Mapping}

The OTF mode minimizes relative calibration errors and pointing errors
across a map.  In this mode, the actual telescope encoder positions are 
read out every 0.01 seconds and folded into the spectra, 
which are read out every 0.1 seconds.   
We set up a typical OTF map to have a total length of 8\arcmin\ on a side, 
which includes a 1\arcmin\ ramp-up and 1\arcmin\ ramp-down distance on either
side of the 6\arcmin\ heart of the map, and we alternated scanning
in the RA and DEC directions for successive maps in order to minimize
striping \citep[cf][]{Em88}.  
We spent about 30 seconds on each map row and observed a
reference position on the sky at the beginning of every second row,
calibrating often enough (every 5--15 minutes) so that the measured 
system temperatures changed by less than a few percent between calibrations.
The maps were of course greatly
oversampled in the scanning direction; in the orthogonal direction,
we spaced the rows by 18\arcsec, which slightly oversamples the maps
even by the Nyquist criterion (22.4\arcsec\ at 115 GHz at the 12 m).
The output antenna temperatures were recorded as the standard
$T_R^*$ = (ON-OFF)/OFF $\times$ $T_{sys}$, where ON is the power measured
on source and OFF is the power measured at the reference position on
the sky (taken to be the OFF measurement nearest in time to a given
ON).
For galaxies where the detected CO emission was clearly confined to a region
within the OTF map boundaries, we later created a new OFF reference spectrum
from the average of roughly 20--40 individual (0.1 second) spectra on either 
side of each row; although this technique may introduce a systematic, 
low-level error into the map, it also greatly improves the baseline stability 
of the spectra.
Each individual OTF map took $\la$ 20 minutes
to complete, and we obtained anywhere from 10 to 30 OTF
maps to achieve the desired sensitivity;
following \cite{Cornwell93}, 
we tried to spend enough time on the total-power 
measurements to match the signal-to-noise ratio of the 
interferometric observations. In practice, when smoothed to 55\arcsec\
for direct comparison, the BIMA maps tended to have lower absolute noise 
levels than the OTF maps by a factor of about two.

Given reasonably stable observing 
conditions, the relative flux calibration across an individual OTF map
should be very good, so that even if the absolute flux scale drifts from map 
to map, the combined final 12 m map should also have very good pixel-to-pixel
calibration. 
An analysis of well-pointed data over many observing
seasons shows that the repeatability of the flux calibration at the 12 m is
accurate to better than a 1 $\sigma$ uncertainty of 10\%.
However, the pointing at the 12 m is a more challenging problem.  For
the OTF data, we were able to crosscorrelate the 12 m data with the
BIMA data (\S 4.3), and we estimate that the absolute pointing accuracy was
good to $\la$ 5\arcsec\ rms after performing the pointing
crosscorrelation.
We subtracted baselines based on linear fits to the emission-free
channels in the passband, then
weighted and gridded the OTF data to an 18\arcsec\ cell in AIPS.
The rms noise levels in the OTF maps were typically 15--20 mK 
(0.50--0.66 Jy~bm$^{-1}$) per 2 MHz (5.2 km~s$^{-1}$) channel.
To allow for some amount of pointing or other systematic
errors in the OTF maps, we assign an overall 1 $\sigma$
uncertainty of 15\% to the flux calibration of the OTF data.

\subsubsection{ Position Switched Spectra}

We observed the remaining 20 galaxies in a simple position-switched  
mode, where
the telescope tracks a specific position on the source for a given 
integration time (typically 30 seconds), then is pointed to a nearby 
(typically closer than $\pm$30\arcmin\ in azimuth from the source) 
reference position to measure the sky brightness.  We observed each 
galaxy at the positions of the BIMA pointings, so that we obtained
7 spectra for the mosaiced BIMA observations and one for the single-pointing
BIMA observations.  Again, spectra of $T_R^*$ were recorded as described
above.  We fitted and removed linear baselines from the spectra using
the Bell Labs spectral line reduction package, COMB.  
The rms noise levels of the spectra were
$\sim$ 5--20 mK per 2 MHz (5.2 km~s$^{-1}$) channel.  

As with the
OTF data, the absolute flux calibration is presumably accurate to
a 1 $\sigma$ uncertainty of 10\%, though this is difficult to
measure directly because we  
were not able to correct for pointing errors, which may be as large
as $\sim$ 15--20\arcsec.  The effect of mispointed data is essentially
unpredictable without an independent way of characterizing it (as with
the BIMA-12 m pointing crosscorrelation for OTF data described in
\S 4.3).  For example, a pointing error of 20\arcsec\ will cause
the measured flux density of a point source to be attenuated by 12\%. 
However, for sources with more complicated structures, the flux density
measured at the mispointed position could be higher or lower than 
the flux density from the desired position.
Furthermore, the measured flux density is a function of the 
observed velocity,
so that the spectral line shape may also be in error.  We therefore
assign a 1 $\sigma$ uncertainty of 25\% to the
flux calibration of the position-switched spectra, keeping in mind that this 
error is dominated by systematic uncertainty.

The position-switched data from the 12 m telescope are summarized in
Table 3, which lists the integrated intensities measured at
each position for a given source.

\section{ Data Reduction and Calibration}

A description of the basic data reduction and calibration is given
in considerable detail in Paper I.  Here, we give a brief overview
and detail a few points that were not discussed more fully earlier.

\subsection{ BIMA Data Reduction }

We reduced the BIMA data using the MIRIAD package \citep{Sault95}.
We removed the instrumental and atmospheric phase variations from
the source visibilities by referencing the phase to observations of
a nearby quasar every 30 minutes.  The antenna-based frequency dependence
of the amplitudes and phases were measured and removed 
using the BIMA online passband calibration at the time of each track.  
Using measurements of the rms phase variation over a fixed, 100 m
baseline \citep{Lay99}, we applied a zeroth-order correction for the 
atmospheric phase decorrelation (Paper I).  The main effect of this 
correction is to create a dirty beam that more accurately represents the 
atmospheric-limited response of the interferometer. 
We set the amplitude scale of the BIMA observations using Mars and Uranus
as primary flux calibrators.   Following Paper I, we assign a 1 $\sigma$
uncertainty to the amplitude calibration of 15\%.

We weighted the visibility data by the noise variance and also applied
robust weighting \citep{Briggs95} to produce data cubes in right
ascension, declination, and LSR velocity.  We included shadowed data
down to projected baselines of 5 m, after checking for false
fringes in the shadowed data; this allowed better measurements of
large-scale emission with the modest penalty of $\la$ 1\% attenuation
of the measured source amplitude.
For most sources, we produced
data cubes with a velocity resolution of 10 km~s$^{-1}$, though we
used 5 km~s$^{-1}$ for the narrow-line galaxy NGC 0628, and 20 km~s$^{-1}$
for NGC 3031, NGC 4725, and NGC 5033. 

For purposes of uniform data reduction, we treated all observations as
mosaics, with the 11/44 single-pointing observations
a trivial subset of the mosaiced case.  Following \citet{Sault99},
we briefly describe the mosaicing process we used as implemented in
the MIRIAD package.  First, the standard mapping task INVERT produces a
``dirty'' image of each pointing separately and then combines the
separate dirty maps into a linear mosaic by taking a weighted average
of pixels in the individual pointings, where the weights are chosen
to minimize the theoretical noise and to correct for the primary
beam attenuation of the individual pointings.  (This ``Sault
weighting'' scheme results in a primary beam function that is 
flat across most of the mosaiced image but that is attenuated
at the edges of the mosaic, so that the noise level does not become
excessive there.  For the single-pointing observations, the primary
beam response is taken to be the usual truncated Gaussian with FWHM of
100\arcsec\ at 115 GHz.)  Because the
$uv$ coverages of the pointings are not identical, the synthesized
(``dirty'') beam also differs slightly from pointing to pointing.
Together with the Sault weighting scheme, this means that the
point spread function (PSF) of the mosaiced image is 
position-dependent.  The INVERT
task therefore also produces a cube of beam patterns, one for
each pointing, so that the deconvolution can compute the
true PSF at any position in the dirty image.
We deconvolved the image cubes using a Steer-Dewdney-Ito CLEAN algorithm
\citep{SDI84}; see Appendix A for a discussion of alternative
deconvolution schemes and issues concerning the deconvolution.   
We did not set any CLEAN boxes so as not to introduce any
{\em a priori} bias as to where we expected emission to be
detected.   We allowed the deconvolution to clean deeply,
to a cutoff of 1.5 times the theoretical rms in each pixel
(see Appendix A, final two paragraphs).
The deconvolved images were restored in the usual way by convolving the
CLEAN solution with a Gaussian with major and minor axes fitted to the
central lobe of one plane of the synthesized beam, then adding the residuals
from the CLEAN solution to the result.  

We performed an iterative, phase-only
multi-channel self-calibration on most detected sources.  We checked that 
the source position was not affected by self-calibration by 
cross-correlating the final BIMA-only map with the {\it a priori}
phase-calibrated map; in most cases, the source moved by $\la$ 0.1\arcsec\
after self-calibration.  Instead, the absolute registration of the maps
are limited by uncertainties in the phase calibration.
Test measurements of two point sources separated by $\sim$ 11\arcdeg\ --
a typical separation for a source and phase calibrator for BIMA SONG
observations -- showed that the absolute registration of the maps should 
be accurate to $\sim$ 0.4\arcsec.

\subsection{ Continuum Emission}

\begin{figure}
\centerline{\includegraphics[height=6.5in,angle=-90]{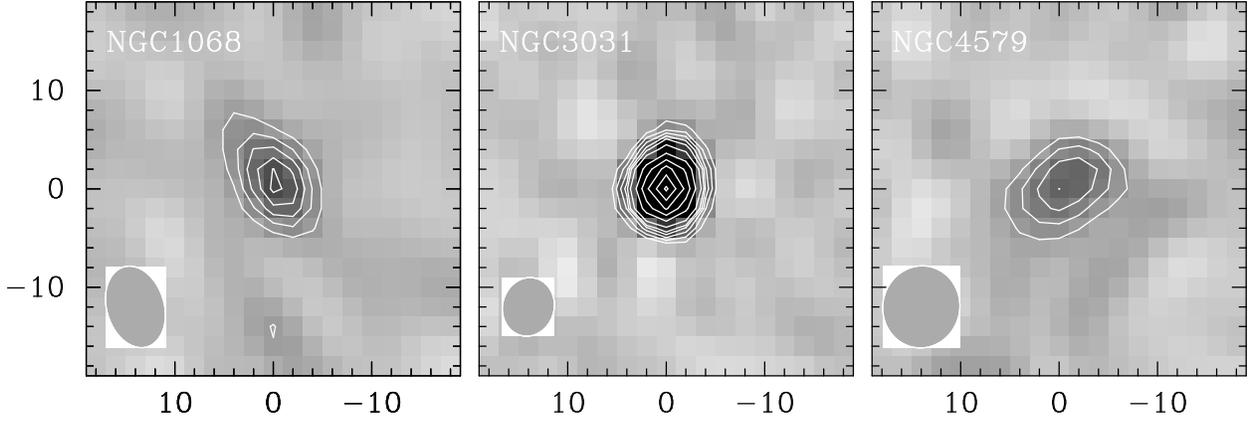} }
\caption{Continuum emission at 112 GHz 
from NGC 1068, NGC 3031 (M 81), and NGC 4579.
Contours are shown at $\pm$2, $\pm$3, $\pm$4, 5, 6, 8, 12, 16, 20, 24, 28
times the measured rms flux density level in the map, which was 6.5 mJy/bm for
NGC 1068, 5.6 mJy/bm for NGC 3031, and 5.6 mJy/bm for NGC4579.  The synthesized
beams, shown in the lower left corner, are 8.5\arcsec$\times$5.8\arcsec\
(NGC 1068), 6.0\arcsec$\times$5.2\arcsec\ (NGC 3031), and 
8.4\arcsec$\times$7.8\arcsec\ (NGC4579).  The maps are centered on the
tracking center listed in Table 1.  NGC 1068 and NGC 4579 are marginally
resolved, while NGC 3031 is consistent with a point source (see \S 4.2).
}
\label{continuum}
\end{figure}

We detected continuum emission at 112 GHz (the lower sideband of the 
local oscillator) in three sources:  NGC 1068 (M 77), 
NGC 3031 (M 81), and NGC 4579.
For all three, we subtracted the continuum from the spectral
line data in the visibility plane by fitting a constant to the real and
imaginary parts of those channels in the passband that were free of
line emission.  We list the 
peak flux density levels of these three galaxies in Table 2 along with 
3 $\sigma$ upper limits for the remaining sources; the maps of the
detected galaxies are shown in Figure \ref{continuum}.
We tested to see whether any of the emission was resolved by fitting 
Gaussians to the three detections.  
The emission from NGC 1068 was marginally resolved; 
the deconvolved continuum source size
was 7.8\arcsec\ $\times$ 2.6\arcsec\ along position angle 34\arcdeg, and
the integrated source flux density was 59.9 $\pm$ 8.1 (formal) 
$\pm$ 9.0 (systematic) mJy.  
These results are consistent with previous measurements
of the 3 mm continuum in NGC 1068, which is resolved into two components,
each of about 40 mJy and separated by about 5\arcsec, one to the
northeast of the other \citep[e.g.][]{HB95}.
The emission from NGC 3031, which is known to have a bright ($\sim$ 90 mJy), 
variable, flat-spectrum radio core \citep[e.g.][]{Crane76, deBruyn76},
was consistent with that from a point source with a flux density of 
161 $\pm$ 6 (formal) $\pm$ 24 (systematic) mJy.  The emission from 
NGC 4579, which like NGC 1068
contains a Sy 2 nucleus, appears to be resolved, with a deconvolved source 
size of 8.7\arcsec\ $\times$ 5.5\arcsec\ along position angle -69\arcdeg\
and an integrated source flux density of 45.0 $\pm$ 7.5 (formal)
$\pm$ 6.8 (systematic) mJy.

\subsection{ Pointing Crosscorrelation}

The implementation of OTF mapping at the NRAO 12 m (\S 3.2.1)
ensures that the internal pointing consistency of an
individual OTF map (taken over $\sim$20 min) is excellent, even if the
overall registration of the map is uncertain.  We were further able to
cross-correlate either the individual OTF maps or averages of several
maps with each other to track slow pointing drifts, allowing OTF maps
taken over many hours, days, or even in different observing seasons, to
have good internal pointing accuracy.  

We performed the cross-correlation using the convolution theorem:
the cross-correlation function of the two maps is the
inverse Fourier transform of the product of the Fourier transforms
of the two input maps.  Since the input maps had many channels,
we summed the resulting cross-correlation function over all channels 
and fitted the position of the peak using a two-dimensional gaussian. 
Using the sum of the cross-correlation function ensures that those
channels with the highest values for the cross-correlation
are weighted the most.

\begin{figure}
\centerline{\includegraphics[height=6.5in,angle=-90]{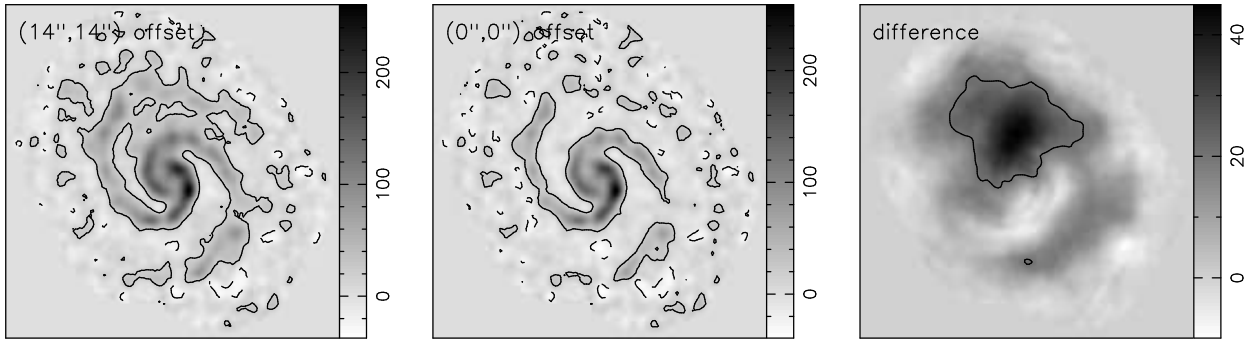} }
\caption{Effect of single-dish pointing error in NGC 5194.  
The combined BIMA+12m maps are shown before ({\it left}) 
and after ({\it middle}) correcting a simulated 20\arcsec\
pointing offset between the 12 m and BIMA maps.  The difference between 
the uncorrected and corrected maps ({\it right}) emphasizes the 
symmetrically placed positive 
and negative errors in the combined map.  Contours are shown at $\pm$ 20 
Jy~bm$^{-1}$~km~s$^{-1}$ to emphasize the difference in the maps.  
The exact flux density distribution of the difference map depends on the 
direction of the pointing error and the source distribution.
}
\label{offset}
\end{figure}

The absolute registration of
the 12 m maps was accomplished by cross-correlating the well-pointed
average 12 m map with the BIMA map, where the BIMA map was smoothed to
match the 55\arcsec\ resolution of the 12 m at 115 GHz.  In performing
this crosscorrelation, we had to assume that the 
extended emission estimated by the deconvolution but not
sampled by the interferometer (Appendix A) -- that is, the
``resolved-out'' flux in the interferometer maps  --
correctly placed the centroid of the emission.
A theoretically more robust method, comparing flux density just in the
region of $uv$ overlap, was limited by poor S/N in the overlap
region of the 12 m data.  In general,
the resulting pointing corrections were 5-10\arcsec, but some were as large as
$\sim$20\arcsec.   If a pointing offset of this magnitude is left
uncorrected, it has the effect of inserting a phase shift into
the combined map, so that there will be symmetrically placed
positive and negative residuals in the final combined map.  An
example of this is shown in Figure \ref{offset}, which shows a
combined BIMA+12m map of NGC~5194 before ({\it left}) and after 
({\it middle}) correcting a simulated 20\arcsec\ offset between the 12 m and
BIMA maps.  The difference between the uncorrected and corrected
maps ({\it right}) emphasizes the symmetric positive and
negative errors introduced in the combined map.  The peak error in
the difference map is nearly 20\% of the peak value in the
correct map of the galaxy.

\subsection{ BIMA and Single Dish Data Combination }

For the 24 galaxies with complete OTF mapping from the NRAO 12 m (Table 2,
\S 3.2.1),
we combined the BIMA data with the single dish data so that
the large-scale emission is accurately imaged.  In Appendix B, we
discuss different techniques for combining interferometric data
with single dish data.  For this paper, we used the combination method 
described in \citet{Stan99} (with a CLEAN-based deconvolution instead of their
MEM).  First, we created a new ``dirty'' (undeconvolved) map 
using a linear combination of the BIMA ``dirty'' map and the OTF map,
where the OTF map was tapered by the BIMA primary beam function
and resampled onto the same $\alpha,\delta,v_{LSR}$ grid as the
BIMA dirty map.
For the linear combination, we chose to weight the images in
inverse proportion to the beam areas.  (In principle, one could
choose any number of weighting schemes.  In practice, however,
putting both images on a common temperature
scale is the appropriate choice in the relatively low S/N regime of 
millimeter data, where there may be a significant contribution to
the total flux in the residual map; see discussion in Appendix A.)
We also created a new ``dirty'' beam using the same weights for
a linear combination of the BIMA synthesized beam and the 12 m beam, 
where the 12 m beam was assumed to be a truncated Gaussian.
We deconvolved the new ``dirty'' map by the new ``dirty''
beam using a Steer-Dewdney-Ito CLEAN algorithm as described in
\S 4.1.
As with the BIMA-only maps, we restored the final images by convolving
the CLEAN solution with a Gaussian (fitted to the central lobe of 
one plane of the combined ``dirty'' beam cube) and then adding the 
residuals from the CLEAN solution to the result.

\section{ A Catalog of BIMA SONG Data }

The full catalog of BIMA SONG figures is available online
at \\
http://astro.berkeley.edu/\~\ thelfer/bimasong\_supplement.pdf 
and through the online edition of the Astrophysical Journal at \\
http://www.journals.uchicago.edu/ApJ/journal.
For each source, the catalog shows the following:

1. ({\it Upper left panel}): The distribution of CO integrated 
intensity (``moment-0'')
emission is shown by red contours overlaid
on an optical or infrared image of the galaxy.  
The integrated intensity maps were formed in the following way.
First, we created a mask by smoothing the data cube by a Gaussian 
with FWHM = 20\arcsec\ and then
accepting all pixels in each channel map where the emission was
detected at a level above 3 times the measured rms 
noise level of the smoothed cube.
Second, we summed all the emission in the full-resolution
data cube after accepting only those pixels that were nonzero in the
mask file.  Finally, we multiplied by the velocity width of an
individual channel to generate a map of integrated intensity in
the units of Jy~bm$^{-1}$~km~s$^{-1}$.
This smooth-and-mask technique is very effective at showing low-level 
emission which is distributed in a similar way to the brighter emission 
in the map.  Such low-level distributions are common, since the deconvolution
often leaves systematic, positive residuals in the
shape of the source (Appendix A), which is 
almost certainly true emission associated
with the source.  Another advantage of the smooth-and-mask
technique is that it does not bias the noise statistics of
the final image, unlike the commonly used method of clipping
out low-level pixels in the channel maps.
However, the smooth-and-mask method may introduce a bias
against any possible compact, faint emission that is distributed
differently than the brightest emission.  

For each map in the upper left panel, the contours are spaced 
logarithmically in
units of one magnitude, or with levels at 2.51 times the previous level.
The starting contour is given in Table 2.  The solid
blue contour denotes the gain=0.5 contour of the primary beam.  For
the single-pointing observations (Table 2), the primary beam response
is unity only at the center of the map and falls off as a Gaussian function
at distances away from the center.  For multiple pointing observations,
on the other hand, each galaxy has a uniform primary beam function
inside a large region bounded by a dashed blue contour;  the gain falls from
1 to 0.5 between the dashed and solid blue contours.

2. ({\it Upper middle panel}): The same CO integrated intensity 
emission shown in the upper left panel is shown as black contours
overlaid on a false-color representation of the CO emission.  The
contours are exponentially spaced at half-magnitude intervals,
or 1.59 times the previous interval.
The starting contour is given in Table 2.  
The false-color minimum is set to the value of the lowest pixel in
the map, which is on average -2.5 Jy~bm$^{-1}$~km~s$^{-1}$.  The
false-color maximum for all maps is 315 Jy~bm$^{-1}$~km~s$^{-1}$,
so that the reader can easily determine the relative strengths of
the flux density for the different sources.
The FWHM of the synthesized beam is shown near the lower left corner of
this box, next to a vertical bar that shows the angular size of 1 kpc at 
the assumed distance to the source.

3. ({\it Upper right panel}): A representative CO velocity field 
is shown in false color.  The fields were generated by
taking the mean velocity at each pixel from Gaussian fits as follows.  
First, a single Gaussian was fit to the spectrum
at each pixel in the datacube.  The initial estimates for the amplitude and 
velocity of the Gaussian were taken from the peak of the spectrum at
each pixel.  The initial 
estimate for the FWHM of the Gaussian was taken to be 30 km~s$^{-1}$
or 60 km~s$^{-1}$ (the fitting routine was run for both values, 
and the results for 60 km~s$^{-1}$ were used when the 30 km~s$^{-1}$ 
fit failed).  The Gaussian fits were
rejected if the velocity error exceeded 2 channels  or
if the integrated flux density of the Gaussian was less 
than $6 \sigma \Delta v$, where $\sigma$ is the rms noise level in
a channel map and $\Delta v$ is the velocity width of a channel.
Finally, the same spatial mask used for the integrated intensity 
images (described in ``upper left panel''
above) was applied to the velocity fields.
The wedge at the right of each velocity field shows the range and
values of the velocities for each galaxy.  Velocity contours are
overlaid on the false-color images; the contours are spaced at
20 km~s$^{-1}$ intervals, except for the narrow-line galaxies
IC~342, NGC~0628, and NGC~3184, which are spaced at 10 km~s$^{-1}$ intervals,
and the wide-line galaxies NGC~3368, NGC~4258, NGC~5005, NGC~5033,
and NGC~7331, which are spaced at 40 km~s$^{-1}$ intervals.

4. ({\it Lower left panel}): The individual channel maps are shown over
the velocity range of emission.   The BIMA SONG data are shown as a
false-color halftone, the stretch of which is shown in a wedge
to the right of the last panel and which ranges from $+2\sigma$ to
0.75 times the peak value in the channel map.  The gain=0.5 contour
is shown as a solid blue contour in the upper left panel.
The LSR velocities are given for the first, second, and last
panels of the map; the velocity width of each channel may be inferred 
from the difference between the velocities listed in the first
and second panels.  Where present, light gray contours show emission
measured in OTF mapping at the NRAO 12 m telescope.  The contours
start at 3 $\sigma_{12 m}$, where $\sigma_{12 m}$ is the measured
rms noise in the 12 m channel map (listed in Table 2), and are spaced 
by factors of 1.5.

5. ({\it Lower right panel}): Spectra comparing the BIMA SONG maps at 
55\arcsec\ resolution are shown along with corresponding 55\arcsec\
spectra from the NRAO 12 m telescope.  For those galaxies mapped
with OTF mapping at the 12 m, a grid of spectra is shown overlaid
on a grayscale representation of the integrated intensity map.
The spectra are spaced by 30\arcsec\, centered on the tracking
center of the source as listed in Table 1; the location of each
spectrum is marked by a light blue cross. 
The red spectrum is the BIMA-only data, smoothed to 55\arcsec;
the blue is the BIMA+12m combined data, smoothed to 55\arcsec,
and the green is the 12 m-only data, at its full resolution of
55\arcsec.  The LSR velocity and flux density scales are shown in a frame
around the lower left spectrum.  For those galaxies not mapped
with OTF mapping at the 12 m, we show instead individual spectra
taken at each of the pointing centers of the BIMA mosaic,
at (0\arcsec,0\arcsec), ($\pm$44\arcsec,0\arcsec), and
($\pm$22\arcsec,$\pm$38\arcsec).  Again,
the red spectrum is the BIMA-only data, smoothed to 55\arcsec,
and the green spectrum is the 12 m-only data.

\section{ Detection Rate }

Since neither CO nor infrared flux density were selection criteria
for the SONG sample, it is of interest to see what the
detection rate of the survey was.  Of the 44 galaxies included, 
only three or 7\% were not detected in line emission.  These are 
NGC 3031 (which was detected in the continuum; see \S 4.2)
and NGC 4699, which were also not detected with the NRAO 12 m telescope,
and NGC 3992, which was formally detected in at least one position
with the 12 m (Table 3).  We note that NGC 3031 (M 81) is a large galaxy with
known CO emission at $r \ga$ 120\arcsec, which is outside our
field of view \citep[e.g.][]{Brouillet91}.  
This means that 42/44 = 95\% of the SONG galaxies have known CO emission.

The high detection rate for CO is not unexpected for a
magnitude-limited sample, since optically bright galaxies might be
expected to be brighter at all wavelengths.  Another contributing factor
is likely to be the Hubble type distribution (Figure \ref{typehist}), 
since Hubble types outside the Sa--Sd range were explicitly excluded, 
and there are few
early-type spirals (Sa--Sab) because the sample is also volume-limited and
such galaxies are rarer.  For comparison, \citet{Young95} detected CO
in 96\% of Sc galaxies, 88\% of Sa--Scd galaxies, and 79\% of galaxies
over all Hubble types.  The tendency for CO to be brightest in
intermediate-type galaxies can be understood in terms of galactic
evolution: early type galaxies such as ellipticals and S0 galaxies have
converted most of their gas into stars, whereas late-type galaxies, while
gas rich, are generally low in metallicity \citep{Zaritsky94}, 
and thus less likely to show detectable CO emission.

\section{ Global Fluxes}

The NRAO 12 m OTF data we have collected for 24 SONG sources comprise
the most extensive collection of fully-sampled, two-dimensional
single-dish maps of external galaxies to date.
We have measured the global flux densities directly in these galaxies,
without having to rely on model fitting of data at selected positions
and the extrapolation of these models to larger radii.  

For 19/24 galaxies, we mapped a core square region of 
6\arcmin$\times$6\arcmin\ with uniform sensitivity; an additional 
1\arcmin\ which was included for a ramp-up or ramp-down distance
extends the total region mapped to 8\arcmin$\times$8\arcmin, though
the outer 1\arcmin\ (square) annulus has significantly lower sensitivity
than the core region of the map.  For NGC 3938, NGC 5033, and
NGC 5247, the core region was 5\arcmin$\times$5\arcmin;
for NGC 5457, the core region was 7\arcmin$\times$7\arcmin;
and for NGC 5194, the core region was 8\arcmin$\times$12\arcmin.

\begin{figure}
\centerline{\includegraphics[height=6.5in,angle=-90]{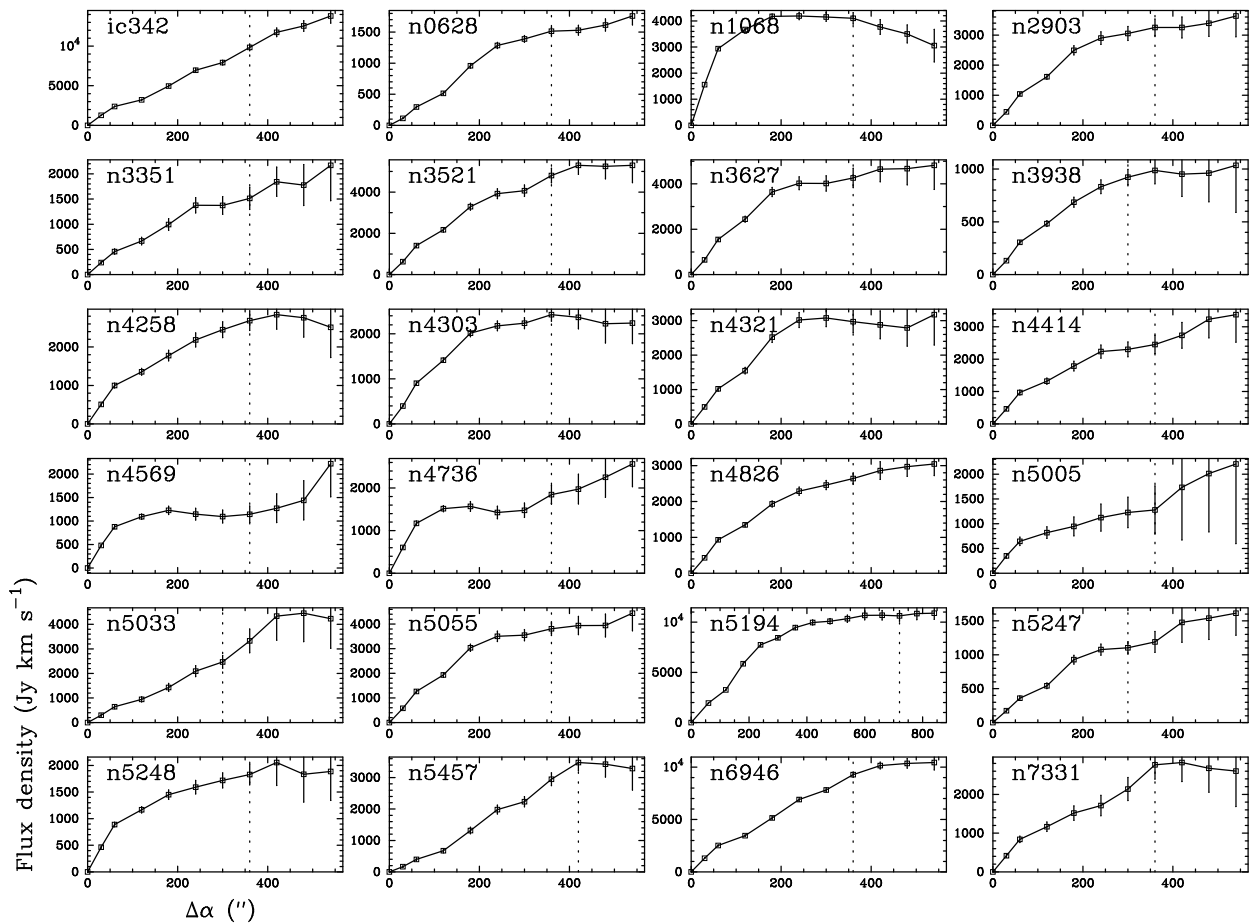} }
\caption{Single-dish total flux densities integrated within a square 
aperture of length $\Delta\alpha$.  The vertical dotted lines indicate 
the size of the 
core region mapped with uniform sensitivity;  at larger apertures, the
noise per pixel increases appreciably.
}
\label{boxflux}
\end{figure}

The global flux densities, measured inside the core regions, are 
listed in Table 4.  The integrated flux densities are also shown 
as a function of the length of a square region over which the
flux density was measured in Figure \ref{boxflux}.
For most of the sources, the flux density at the core length 
$\Delta\alpha$ (denoted
by the vertical dotted lines) has achieved a plateau, which
suggests that there is very little emission from CO at larger
regions than measured.
For large sources like IC 342, the flux density is clearly still
increasing at large $\Delta\alpha$; this is consistent with the known
CO distribution of this galaxy, which extends to nearly 
15\arcmin$\times$15\arcmin\ on the sky \citep{Crosthwaite01}.
Some of the sources, like NGC 4569 and NGC 5005, show a large 
increase in
flux density outside the core region even though the emission
appears to have achieved a plateau inside the core region.
While it may be that the flux density for these sources truly 
increases outside
the core region, the error bars on the points at large $\Delta\alpha$
are large enough that the measured points are also consistent
with the flux density measured at the core length.

\begin{figure}
\centerline{\includegraphics[height=6.5in]{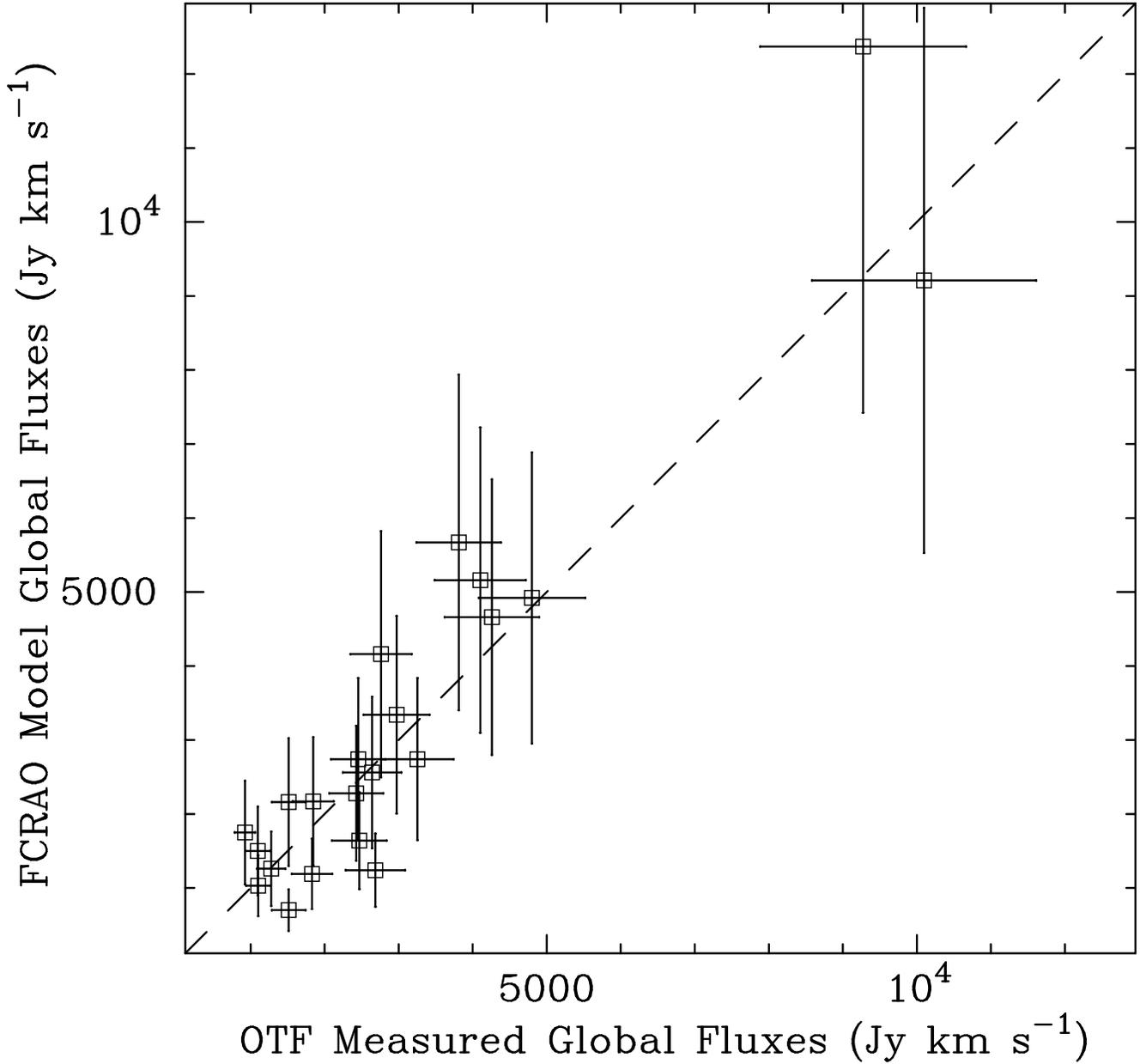} }
\caption{Comparison of global flux densities measured from OTF data with
modeled global flux densities from the FCRAO Survey \citep{Young95}.
The dashed line indicates equal values on the two axes.  See \S 7 for
discussion.
}
\label{globflux}
\end{figure}

We compare our measured OTF global flux densities with global flux
densities derived from model distributions using the FCRAO 
Survey \citep{Young95} in Figure \ref{globflux}.  In general,
the agreement is very good; the average ratio of the OTF to
FCRAO flux density is 1.05, with a standard deviation of 0.19.
The worst discrepancies are for the galaxies with the lowest
global fluxes, like NGC 3351 and NGC 4258, where the OTF measurements 
are both 2.2 times higher than the FCRAO flux density, or NGC 3938,
where the OTF measurement is 0.53 times the FCRAO flux
density.  This tendency was also apparent when we compared integrated
intensities from the position-switched data from the 12m 
telescope (Table 3) with those at matched positions from the FCRAO
survey:  the weaker sources disagreed by as much as a factor of
$\sim$ 2, while the stronger sources agreed to within about 20\%.

Table 4 also lists the CO luminosity and total CO mass inside
the region measured for each OTF source.  Assuming a constant
CO/H$_2$ ratio of 2 $\times 10^{20}$ cm$^{-2}$ (K km~s$^{-1}$)$^{-1}$,
the calculated masses range from 2.4$\times$10$^8$ M$_{\sun}$ for
NGC 4736 to 7.4$\times$10$^9$ M$_{\sun}$ for NGC 5248; the
average of the measured OTF masses is 3.3$\times$10$^9$ M$_{\sun}$.

\section{ Flux Recovery}

An interferometer is fundamentally limited by the minimum spacing of its
elements.  Because two elements can never be placed closer than some
minimum distance $S_{min}$, signals on the sky larger than some
size $\propto$ $\lambda/S_{min}$ will be severely attenuated.
This effect is commonly referred to as the ``missing flux'' problem.
In this section, we present an analysis of the issue of large-scale
flux recovery as measured using a comparison of BIMA-only data with the
total flux as measured at the NRAO 12 m telescope.  We emphasize that
for the 24 BIMA SONG maps that incorporate 12 m OTF data into them,
the final BIMA SONG maps presented in this paper are not missing
any flux.  Nonetheless, an analysis of the BIMA-only data for these
sources leads to some new results on this technical issue, which are
discussed below.

\subsection{ Imaging Simulations of Large-Scale Flux Recovery}

\begin{figure}
\centerline{\includegraphics[height=5in]{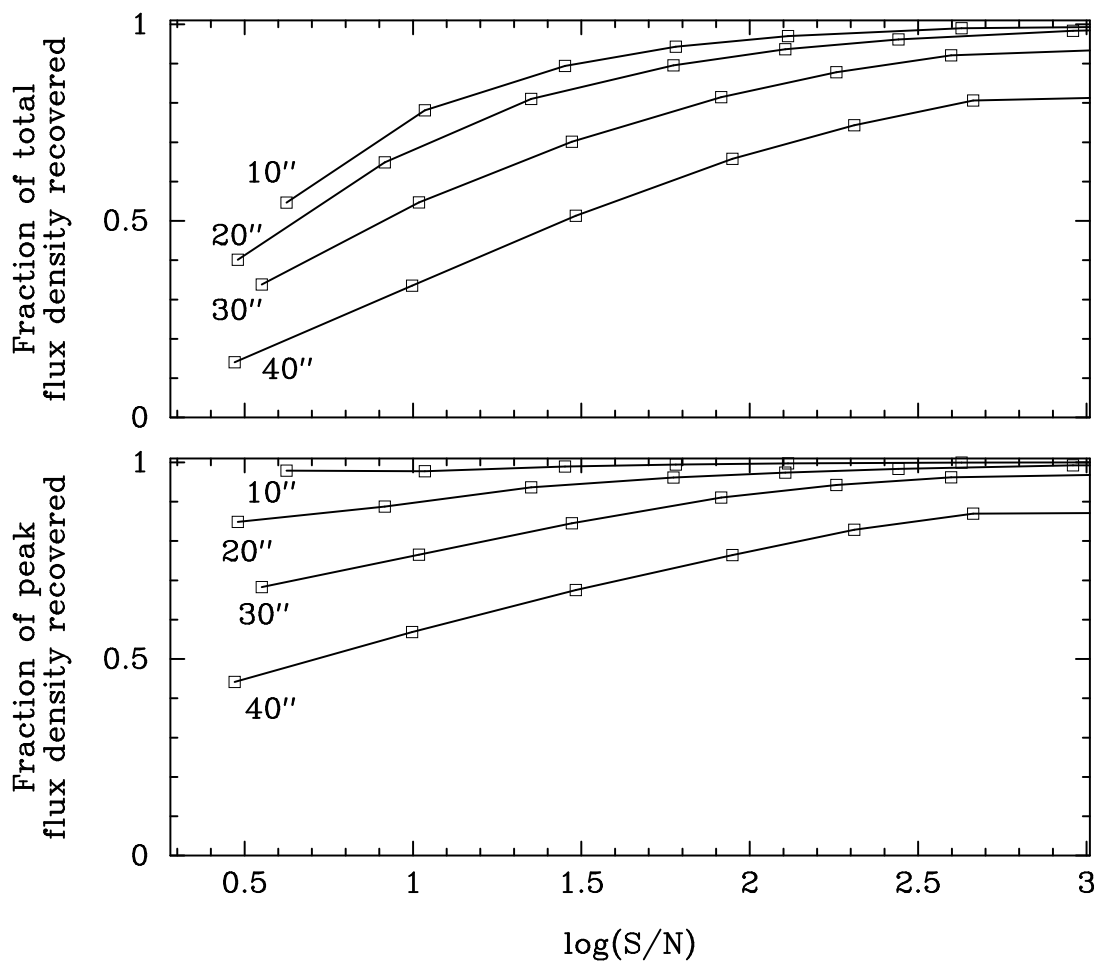} }
\caption{ Fractions of total flux density recovered 
inside a $\pm$100\arcsec\ box 
({\it top}) and peak flux density  recovered in the map ({\it bottom}) for 
BIMA simulations of spiral models as a function of the log of the 
signal-to-noise ratio.  Simulations are shown
for spiral models with arms widths of 10, 20, 30, and 40\arcsec.
The plots reflect the fact that it is increasingly difficult
for the deconvolution to reconstruct large-scale emission
as the signal-to-noise
ratio decreases.  Figure and caption are adapted from \citet{Helfer2002}.
}
\label{sn}
\end{figure}

With the BIMA SONG observations for motivation,
\citet{Helfer2002} recently presented a study of imaging simulations of 
large-scale flux recovery at millimeter 
wavelengths.  These simulations showed that the nonlinear deconvolutions
that are routinely applied to interferometric maps interpolate and
extrapolate to unsampled spatial frequencies and thereby reconstruct
much larger scale structures than would be suggested by an
analytical treatment of the flux recovery problem.  However, 
\citet{Helfer2002} also showed that
fraction of flux density recovered for a given observation is a function
of the S/N ratio of the map.  This is illustrated in Figure \ref{sn},
which shows the fractions of total and peak flux densities as a function of
the S/N ratio for simulated BIMA observations of sources with widths of 
10--40\arcsec.   For S/N ratios of 10--40,
which are typical of individual channels for BIMA SONG maps,
we could expect to reconstruct roughly 80--90\% of the total flux density 
of a 10\arcsec\ source and roughly 30--50\% of the total flux density of a
40\arcsec\ source, whereas for noise-free simulations, BIMA recovers
essentially all of the total flux density of a 10\arcsec\ source and 
80\% of the total flux density of a 40\arcsec\ 
source \citep[see Table 2 in][]{Helfer2002}.  
The fraction of the peak flux density recovered (bottom
panel of Figure \ref{sn}) is typically somewhat higher than the fraction
of total flux density recovered. 

\subsection{ Flux Recovery for BIMA Maps with OTF Data}

In each of the 24 galaxies for which we have complete OTF mapping from the
NRAO 12 m telescope (Table 2), we were able to make a 
pixel-by-pixel measurement
of the flux recovery of the BIMA-only maps.  To do this, we first
divided the BIMA data cube by the appropriate
(mosaiced or non-mosaiced) primary gain 
function; we then smoothed the data cube to the 55\arcsec\ resolution
of the 12 m maps.  Finally we summed over the planes with emission 
in order to make a map of integrated intensity.
We made the corresponding integrated intensity maps of the
12 m emission and then formed ratio maps of the smoothed BIMA-only maps to the
12 m maps.  Finally, we masked the ratio
maps to show only those pixels that are non-zero in the masked, clipped moment
maps shown in the BIMA SONG catalog in \S 5, and we clipped outside 
the original gain=0.5 contour
(also shown in the maps in \S 5).  This masking and clipping serves the
purpose of eliminating pixels with the highest formal errors in the
ratio maps.
Given the assigned 1 $\sigma$ uncertainty of 15\% for the flux
density in the BIMA maps and 15\% for the 12 m 
maps, the 1 $\sigma$ error in the ratio maps is 21\%.  

\begin{figure}
\centerline{\includegraphics[height=7.0in]{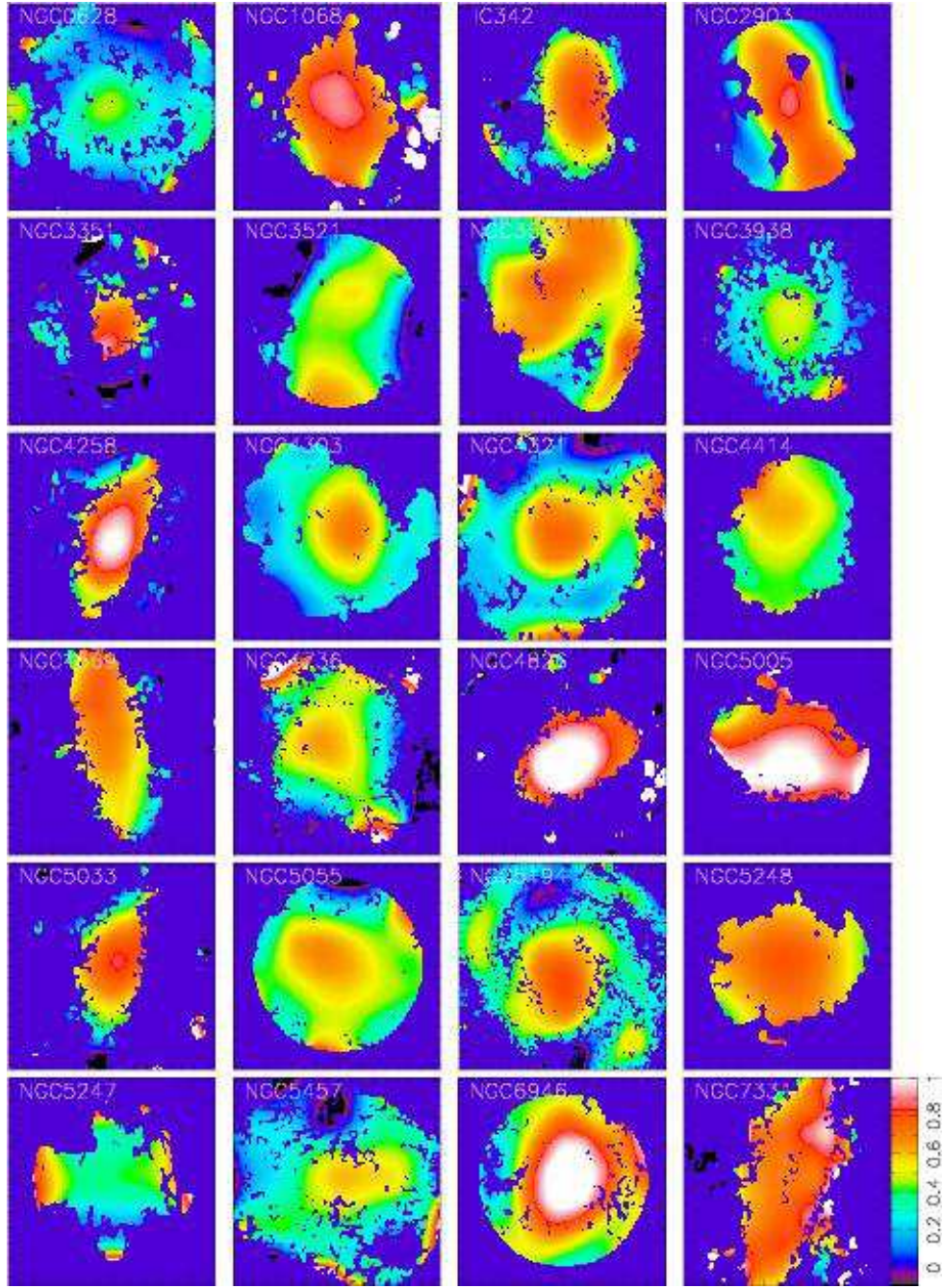}}
\caption{ Flux recovery ratio maps of the BIMA-only data, smoothed to 
55\arcsec, compared with the 12 m data.  The scale is shown as a wedge in the
lower right panel; a value of unity indicates that all of the single-dish
flux was recovered in the BIMA-only  map at that location.  
The 1 $\sigma$ uncertainty in the ratios is about 20\%.
}
\label{recov}
\end{figure}

The flux recovery ratio maps are shown in Figure \ref{recov}.  
We can see from the figure
that there is a remarkable variety in the flux recovery from source to source:
for NGC~1068, NGC~4258, NGC~4826, NGC~5005, and NGC~6946, essentially
all of the single-dish flux density is recovered by the BIMA maps, at least in
the regions of brightest emission.  At the other extreme, for 
NGC~0628, NGC~3938, and NGC~5247, only 30--50\% of the total single-dish 
flux density is recovered over most of the strongest emitting regions. 
The remaining 16 sources recover intermediate amounts of flux density, with
values typically about 70--90\% over the brightest regions.
Clearly, the ability of any given interferometer to reconstruct 
structures in a source is very dependent on the source itself.

It is clear from Figure \ref{recov} that the fraction 
of flux density recovered 
by a given telescope is a function of location 
within the map.  Even
for a source like NGC~6946, which has a dominant, centrally concentrated
region of emission in the nuclear region 
which is fully recovered by the interferometer (Figure \ref{recov}), 
we can see that away from the nucleus, the fraction of flux density recovered 
falls to 30--40\%.  This is also apparent in the grid of spectra 
shown in the Catalog figure for NGC 6946.
This trend is generally apparent (if we exclude galaxies with
ring-like emission, like NGC~3521 and NGC~7331):  the farther from the
nucleus, the less the fraction of flux density recovered.  We can attribute 
this trend to three factors:  first, for any given velocity in a data cube,
there is a larger region of sky available to emit at locations away from 
the center of the galaxy rather than at the center (that is, the 
velocity gradient is largest at the nucleus, which tends to confine
emission near the nucleus to small widths in individual channel maps).  
Second, the tendency
is for the brightest emission and best sensitivity to be near the 
center, so that the
S/N ratio is highest in the central region.  
From Figure \ref{sn} and the description of imaging simulations above
(\S 8.1), we would expect the deconvolution to reconstruct less
of the total flux in regions of lower S/N.
Third, for the BIMA SONG mosaiced galaxies, only the central pointing
of the 7-pointing observations
had the full benefit the \citet{EkersRots79} effect.
\citet{EkersRots79} argued that each $uv$ point measured included contributions
from effective baselines as short as $S_{min}-D$, where $S_{min}$ is the
minimum (projected) center-to-center separation of the dishes and 
$D$ is the dish diameter, and they showed that one could effectively
extend the $uv$ sampling inwards to $S_{min}-D$ by scanning
the telescope over the source.  However, for the 26-field mosaic
of NGC 5194 as well as the single-pointing observations of
NGC 4414 and NGC 5005 (and to a lesser extent, NGC 5248 and NGC 5247),
the fraction of flux density recovered still falls as a function of
distance from the nucleus.  We may conclude that the lack of
the required Ekers and Rots sampling for the outer fields of the
BIMA SONG mosaics
is somewhat less of a valid explanation for the observed trend
than the other two factors described above.

\begin{figure}
\centerline{\includegraphics[height=6.5in,angle=-90]{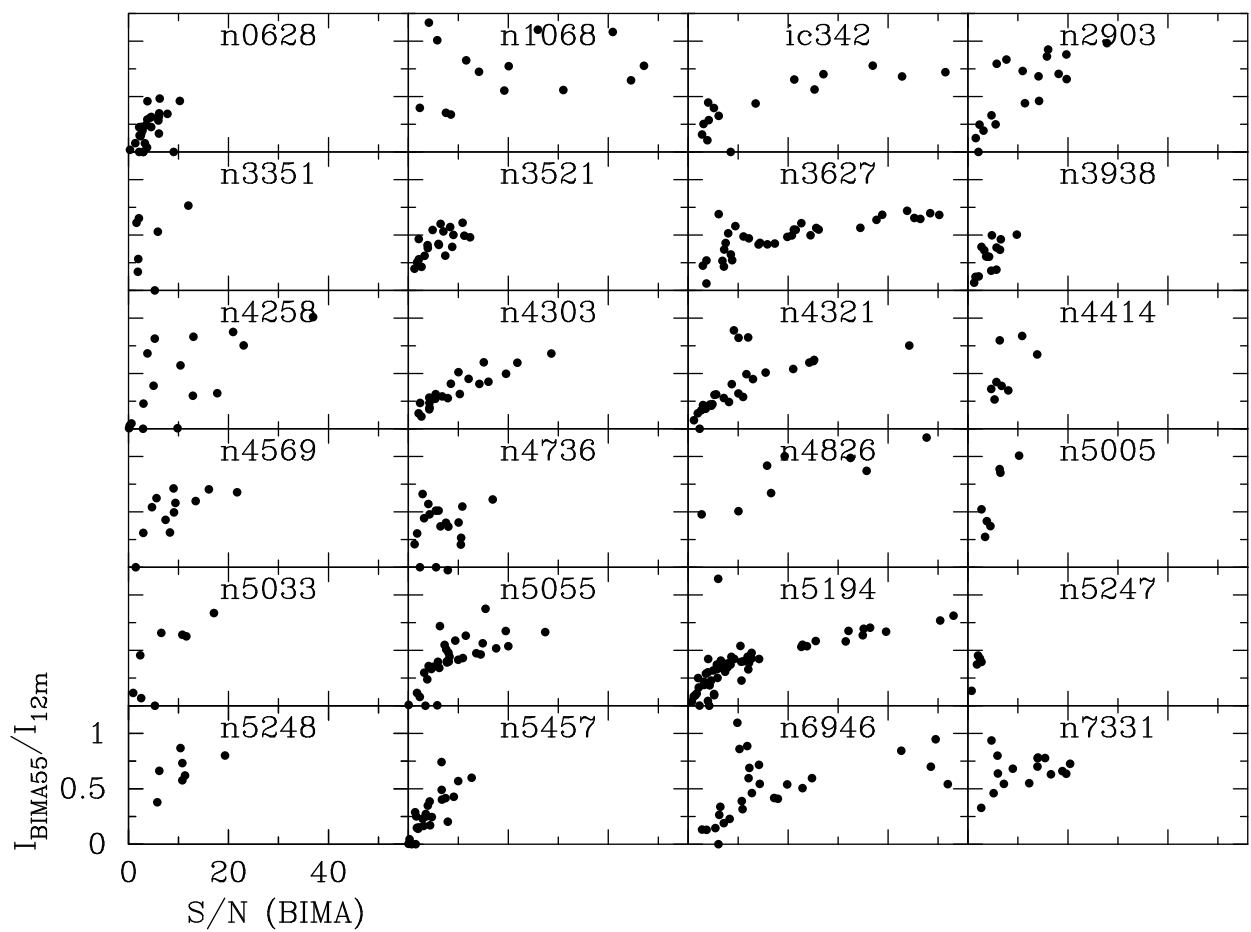} }
\caption{ Ratio of flux recovery as a function of S/N ratio of the
unclipped BIMA maps of integrated intensity.  To form the flux
recovery ratio, the BIMA-only maps were smoothed to 55\arcsec\ and divided
by the OTF maps from the NRAO 12 m telescope.  
Each dot represents a half-beamwidth (27.5\arcsec) sample.
Note the general tendency
for a larger ratio of flux recovery as the S/N ratio increases for
a given source.
}
\label{sndata}
\end{figure}

The dependence of the flux recovery on the S/N ratio of the interferometric
data predicted by the simulations is readily apparent in real data.
Figure \ref{sndata} shows the
ratio of total flux recovered as a function of the S/N ratio for each
of the 24 galaxies with OTF maps.  With the possible
exception of NGC 1068, each source clearly shows the predicted trend.
(From inspection of the ratio map of NGC 1068 in Figure \ref{recov}, 
those pixels with the highest measured ratios probably have the highest
formal errors in the ratio as well.)

For the 24 galaxies shown in Figure \ref{recov}, we have of course
incorporated the 12 m OTF data in the BIMA SONG maps so that the
total flux density is accurate for these maps.  Furthermore, structures larger
than 55\arcsec\ should be accurately represented in the BIMA+12m SONG maps,
and structures $\la$ 20\arcsec\ should also be accurately represented in
the maps.  It is difficult to be confident of the accuracy of structures 
on the intermediate size scales of 20--55\arcsec.  In all likelihood,
these structures are somewhat misrepresented as their attenuated BIMA-only
structures sitting atop a smoothed-out 12 m plateau.

\subsection{ Flux Recovery for BIMA Maps with PS Data}

The remaining 20 BIMA-only maps typically have lower S/N
ratios than the BIMA+12m maps; this is because CO brightness was a
selection criterion in deciding which sources to observe with
OTF mapping at the NRAO 12 m.  These galaxies were instead
observed in a simple position-switched mode at the 12 m.
Given the lower S/N ratios, we
would expect to reconstruct less flux density than for the brighter sources
shown above \citep[\S8.1;][]{Helfer2002}. 

For the BIMA-only galaxies, we compared integrated intensities of spectra taken 
with the NRAO 12 m telescope at each of the BIMA pointing centers with 
the flux density measured in the BIMA-only maps, smoothed to 55\arcsec.  
As with the BIMA+12m maps, in order to compute the ratio of flux density 
recovered, we first divided the BIMA data cube by the (mosaiced or 
non-mosaiced) primary gain.  We then summed those planes in the smoothed
data cube that correspond to the velocity limits of integration in the 
12 m spectra.  The 12 m and 55\arcsec-smoothed BIMA spectra 
are displayed in the BIMA SONG catalog in \S 5; the
ratios of flux density recovered are also listed in the last column of Table 3
along with their formal uncertainties.
Given the assigned 1 $\sigma$ systematic uncertainties of 15\% in the BIMA maps
and 25\% in the 12 m spectra (\S 3.2.2), the systematic 1 $\sigma$ error in 
the ratios are 30\%.

\begin{figure}
\centerline{\includegraphics[height=6.5in,angle=-90]{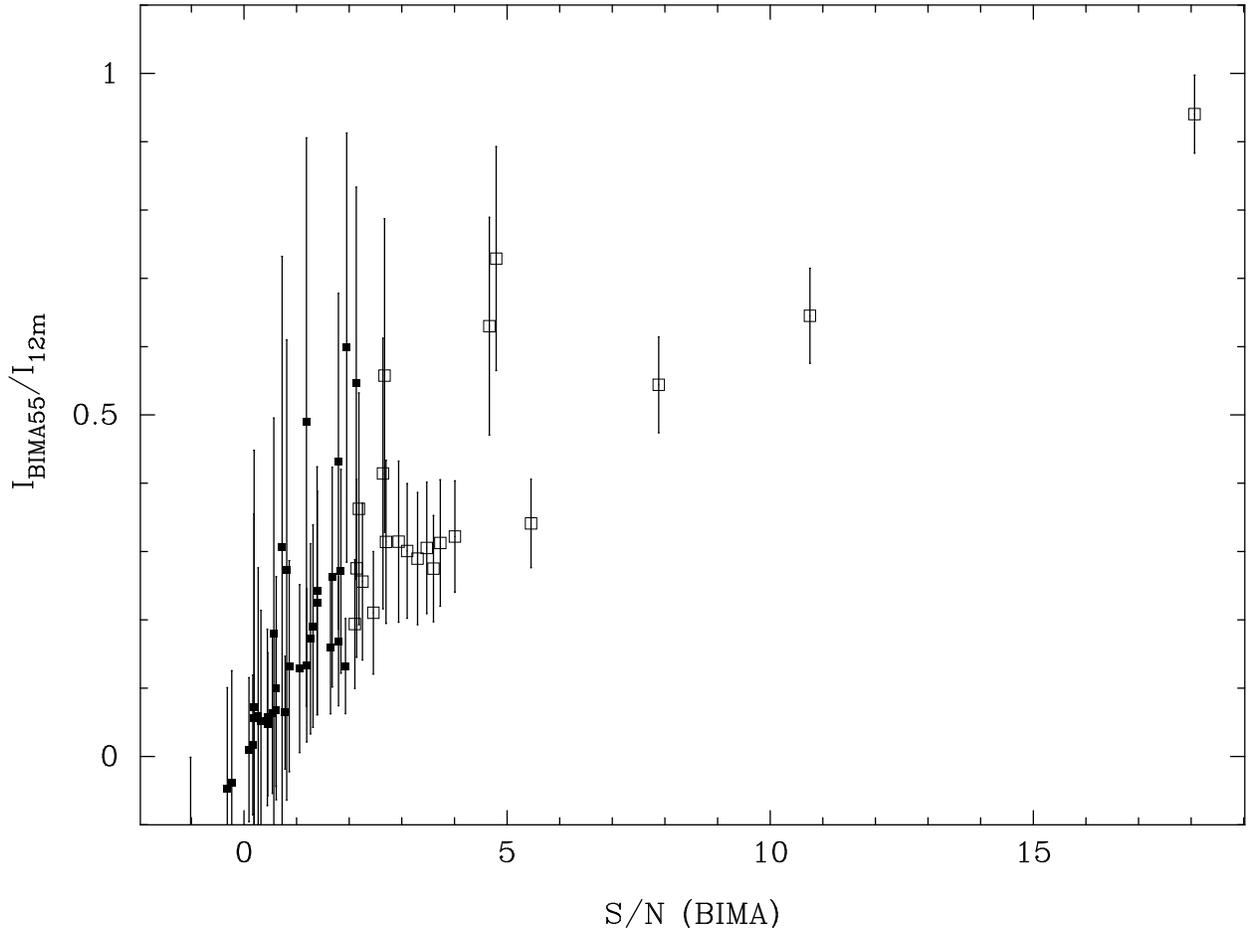} }
\caption{ Ratio of BIMA flux density relative to 12 m flux density,
as a function of the S/N of the unclipped BIMA data.  To form the ratio, the
BIMA-only maps were smoothed to 55\arcsec\ and compared to PS spectra 
from the NRAO 12 m telescope at discrete positions only.  Data from 
55 positions detected
at the 12 m are shown.  The open squares are ratios detected at the $\ge$
2 $\sigma$ confidence level, while the filled squared are ratios measured
at $<$ 2 $\sigma$.  1 $\sigma$ error bars are shown.
}
\label{ratio_sn}
\end{figure}

Of the 97 positions observed at the 12 m and listed in Table 3, we
detected 55 positions.  Of these 55 positions that were detected
at the 12 m, the ratios of the BIMA-55\arcsec\ flux density to 12 m flux density
are measured at 21 positions with a $\ge$ 2 $\sigma$ confidence level.
The ratios for all 55 positions are plotted in Figure \ref{ratio_sn},
which shows the 21 detected ratios as open squares and the remaining
34 ratios as filled squares.  
As the figure shows, there is a clear trend to detect higher ratios
where the S/N of the BIMA data is higher.
Furthermore, as with the BIMA+12m maps, the ratio of flux density recovered
tends to be higher in the central position than in the outer regions
of the mosaic, at least for the brighter galaxies.  
We may conclude that, given better S/N measurements, we would likely
be able to reconstruct more flux density than has been measured here.
Furthermore, following Figure \ref{sn}, we might expect that the peak
flux density that is measured in the BIMA SONG maps is a more reliable
indicator of the flux density in the source than the total flux density 
is.  This is exemplified in the BIMA and 12 m spectra shown for NGC 3184 in 
the BIMA SONG catalog entry (see  \\
http://astro.berkeley.edu/\~\ thelfer/bimasong\_supplement.pdf), 
where the ratio of peak flux density  recovered for the
central region is about 75\%; compare this with the ratio of
total flux density  recovered of 54\% from Table 3.  

From the results of \citet{Helfer2002} and from the above examples,
it is clear that it is very difficult to conclude whether some or
all of the BIMA-only SONG maps are systematically missing large-scale emission,
or whether the SONG maps are instead reasonable representations of
the sources, albeit at low S/N.  It may be reasonable to assume
that those maps of galaxies with only compact structure 
apparent, like NGC 3726 or NGC 4725, may be fairly accurate 
representations of the sources.
Maps of galaxies with clearly extended structures, like NGC 3184 or NGC 4535,
are almost certainly missing some of the larger-scale
flux density.

\section{ Do Molecular Distributions Peak at the Centers of Galaxies?}

In the FCRAO Extragalactic CO Survey \citep{Young95}, only 15\%
of the 193 galaxies observed at multiple positions at 45\arcsec\
resolution showed molecular distributions that lacked a central
peak.  In Paper I, we showed that our single-dish data were
consistent with this result, namely, that all 15 Paper I galaxies
showed single-dish radial profiles that rise monotonically to the
nucleus.  Thus, it appears that the bulk of molecular gas in nearby
spiral galaxies tends to be concentrated within the central $r \la$ 5 kpc.

\begin{figure}
\centerline{\includegraphics[height=6.5in,angle=-90]{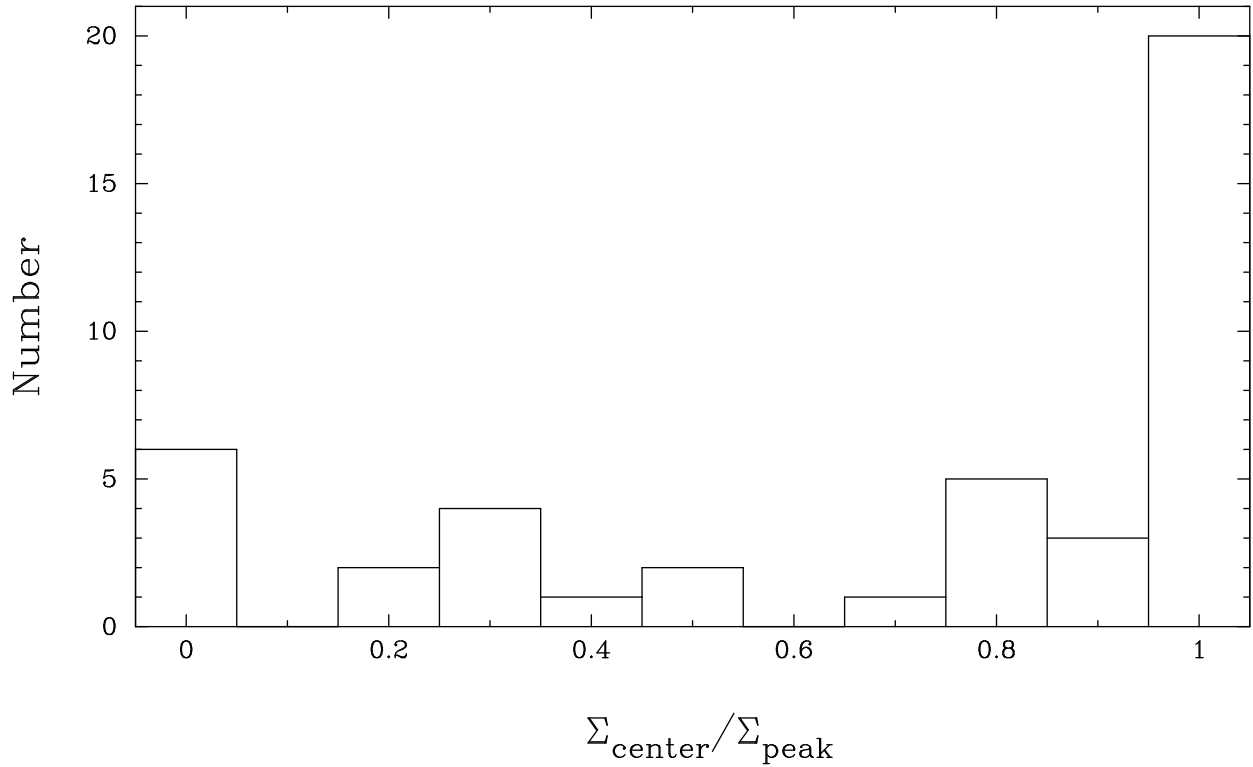} }
\caption{ Number of galaxies exhibiting each fraction of central
peakedness in the molecular surface brightness.
}
\label{centerfrac}
\end{figure}

At 6\arcsec\ resolution, however, about half of the SONG subsample did
not exhibit central peaks (Paper I), a result that was unlikely to
have been caused by differences in the two samples.
For the full sample, we can quantify the degree of the central
concentration by comparing the molecular surface density at the
center of the galaxy to the peak molecular surface density 
in the galaxy.  The ratio of central to peak surface brightness
should be fairly insensitive to the amount of flux resolved out,
so even for the 20/44 maps which lack single-dish data, the ratio
should be a valid measure of the central peakedness.
We list the peak surface density and the central surface density
(taken as the peak within the central 6\arcsec\ beam) in Table 5,
and we show a histogram of the distributions in Figure \ref{centerfrac}.

As Figure \ref{centerfrac} shows, just 20/44 or 45\% of the SONG galaxies
have their peak molecular surface densities within the central 
6\arcsec\ or $\sim$ 360 pc.  At the other extreme, 6/44 or 14\% of
the galaxies lack any detectable molecular gas within the central few hundred
pc.  The remaining 18/44 sources show intermediate
fractions, where there is detected CO at the center but at a level
that is weaker than the peak surface density.   We note that two
of the galaxies with particularly low fractions -- NGC 0628 (0.20) and 
NGC 5194 (M51, 0.28) are grand-design spirals with two dominant spiral arms 
that may be traced all the way to the nuclear region.  The integrated
intensity maps of these galaxies 
suggest a continuity of the large scale structure down to
the 100-pc scales in the nuclear regions of these galaxies, in
contrast to galaxies like the Milky Way (\S 10) or SONG galaxies
like NGC 6946, which show central concentrations in the inner
$\sim$ 500 pc that are an order of magnitude higher in molecular surface
density than are seen in their disks.

\section{Comparison with Local Group Morphologies}

        Galaxies like the Milky Way and M31 produce most of the optical
luminosity in the nearby universe because they have luminosities close
to the knee in the luminosity function (L$_*$) \citep[e.g.][]{BinneyMerri98}. 
It is reasonable then to ask the following: to
what degree are their CO distributions represented in the SONG
sample?  In this section, we present a basic comparison of Local
Group molecular morphologies and brightnesses to those from the survey.
We defer any more complex analysis of underlying causes for the
morphologies, whether caused by bars or other dynamical considerations
\citep{Regan02}, galaxy luminosity, size, or other factors.

We first determine the characteristic sensitivity of the SONG
survey.  The typical rms noise is about 58 mJy in a 6.1\arcsec\ beam
when the velocity channels are averaged to 10 km~s$^{-1}$, a velocity
resolution comparable to the FWHM linewidth of a GMC in the Milky Way
\citep{Blitz93}.  
(The best sensitivity achieved by SONG is a factor of 2.5 lower than this
average; see Table 2.)
Using a CO/H$_2$ conversion factor of 2 $\times 10^{20}$
cm$^{-2}$ (K km~s$^{-1}$)$^{-1}$, this corresponds to an rms column density
sensitivity of 4.6 M$_{\sun}$ pc$^{-2}$ in each 10 km~s$^{-1}$ velocity channel.
Isolated points in the SONG maps are probably not significant at less
than 3$\sigma$ or 13.7  M$_{\sun}$ pc$^{-2}$, but the edges of extended
emission are probably reliable at 2$\sigma$ or 9.1 M$_{\sun}$ pc$^{-2}$.
The typical spatial resolution in  a 6.1\arcsec\ beam is 350 pc at the
median distance of 11.9 Mpc, but is a factor of 5 smaller for the
nearest galaxies.  For comparison, the mean surface density of a GMC in
the solar vicinity is 50 - 100 M$_{\sun}$ pc$^{-2}$ \citep{Blitz93}, within an
area of 2.1 $\times 10^3$ pc$^{2}$.

	The distribution of CO in the Milky Way can be described as
having a central molecular disk, a zone of small but indeterminate
surface density out to about 3 kpc, increasing to a peak at 4 - 7 kpc
and declining monotonically thereafter \citep[e.g.][]{Dame93,Blitz97}.
The inner disk has a radius of about 300 pc and a surface density of
about 500 M$_{\sun}$ pc$^{-2}$ \citep{Guesten89}; this surface
density may be
an overestimate if the CO/H$_2$ conversion factor is low as suggested
by some studies \citep{Sodroski94}, and it is in any event 
accurate only to within a factor of two.  The disk is probably oval in
response to the central bar \citep{Binney91}.  Beyond the central
disk, the surface density declines to a small value, and then increases
abruptly at a distance of about 3 kpc, reaching a broad maximum of
about 6 M$_{\sun}$ pc$^{-2}$ at a distance of about 4 kpc from the center,
thereafter declining approximately exponentially to a distance of about
15 kpc.  Observed at an inclination of 45\arcdeg, the azimuthally averaged
column density at the peak of the distribution in the disk would be
about 8.4 M$_{\sun}$ pc$^{-2}$, which is close to the detection 
limit of our survey.
Since the CO appears to be concentrated in the spiral arms, the local
density might be as much as two or three times this value, and the peak of
the molecular distribution in the disk would be detectable in the SONG
survey.  Thus, the Milky Way would appear in our survey as a bright,
somewhat resolved central disk with a ring of spotty emission at a
distance of 4 - 7 kpc.

	We note first that at the average distance in our sample 
of 12 Mpc,  the inner disk of the Milky Way 
would contain only two or three resolution
elements, and the so-called molecular ring would lie at a radius of
60\arcsec - 90\arcsec.   For galaxies at half that distance, only the
central disk would be observed, with a diameter of about 20\arcsec,
and possibly quite elongated.  Twenty of the 44 galaxies in the SONG have
peak surface densities in excess of 300 M$_{\sun}$ pc$^{-2}$ (Table 5), so the
large central disk in the Milky Way is not at all unusual, even
taking into account the large uncertainty in the mass.  Fewer of the
galaxies have a high surface density central disk along with a large
diameter, low surface density ring, but perhaps the distribution
closest to that of the Milky Way is NGC 3351.  It has a small inner
elongated molecular structure with a peak surface density of 
about 560 M$_{\sun}$ pc$^{-2}$.  
The inner CO is nearly perpendicular to the large-scale
optical bar, suggesting that the gas resides on x$_2$ orbits, similar
to what is inferred for the molecular gas in the bar of the Milky Way
\citep{Binney91}.  Outside the central structure, the CO is absent 
to about 50\arcsec; the CO
becomes just detectable at a distance of about 3 kpc, similar to what
is seen in the Milky Way.  The surface brightness of the molecular
ring in this galaxy seems a bit smaller than that in the Miky Way,
however.  Another galaxy with a similar CO morphology and central CO
disk is NGC 4535.  Its central CO disk has a surface density 
of 610 M$_{\sun}$ pc$^{-2}$, which is similar
to that of NGC 3351, but the gas that makes up its molecular ring has
a somewhat higher surface density than that in the Milky Way.  
The CO distribution of the Milky Way thus appears to be bracketed by
these two galaxies, and is not too dissimilar from several others.

The total mass of molecular gas in the Milky Way is about 1.2 $\times$ 10$^{9}$
M$_{\sun}$ \citep{Dame93,Sodroski94}.  
This is somewhat higher than the median mass of the SONG sources,
5 $\times$ 10$^{8}$ M$_{\sun}$, though it is somewhat lower than the
average of the sample \citep[where we inferred the non-OTF masses from
measurements of 15 sources by][]{Young95}, which is
about 3 $\times$ 10$^9$ M$_{\sun}$.   The masses of NGC 3351 and NGC
4535 are 1 $\times$ 10$^9$ M$_{\sun}$ (Table 5) and 
3 $\times$ 10$^9$ M$_{\sun}$ \citep[from][]{Young95},
which are comparable in magnitude to that of the Milky Way.

	The CO distribution in M31 is quite different.  It has a very
low surface density of molecular gas inside a radius of 6 kpc 
\citep{Loinard96,Loinard99}. 
Like the Milky Way, the CO declines monotonically
beyond 8 kpc with a similar surface density and radial distribution.
However, the peak azimuthally averaged surface density is only about 2
M$_{\sun}$ pc$^{-2}$ and reaches local maxima of about 3 M$_{\sun}$ pc$^{-2}$.
Taking into account the increase due to inclination effects, and
non-axisymmetry, the peak observed column density could rise to as much
as 5 M$_{\sun}$ pc$^{-2}$, but probably not much more.  M31 would thus
probably be undetectable, or at best, be marginally detectable as a
thin, very patchy ring with a radius of 8 kpc, if the inclination
angle were higher than average.  In the SONG survey, even if such a
ring were detectable, it would be observed only at radii beyond
60\arcsec\ in the most distant galaxies.

In terms of general morphology, several galaxies have large
diameter rings, which may be the peaks of exponentially declining
distributions, without central CO concentrations.  These include NGC
2841, NGC 3953, and NGC 7331, though in all of these cases the
ring is smaller ($\sim$ 5 kpc) and has a larger surface density.
Thus the overall CO morphology of M31 is not unusual in our sample.
The closest match to M31 appears to be NGC 3953, with a mean
surface density of about 12 M$_{\sun}$ pc$^{-2}$.  However,
the CO mass of M31 \citep[$\sim$ 2 $\times$ 10$^8$ M$_{\sun}$, 
based on][]{Heyer00} is considerably
lower than that we infer for NGC 3953 based on the FCRAO
survey \citep[4 $\times$ 10$^9$ M$_{\sun}$,][]{Young95}.
Several other distant galaxies have little or no molecular gas observed at our
limiting surface brightness: NGC 3992, NGC 4450, NGC 4548, and NGC 4699.
The total masses we infer from the FCRAO survey for NGC 4450 and
NGC 4548 are $\sim$ 10$^9$ M$_{\sun}$; NGC 3992 and NGC 4699
were not measured at FCRAO.
Some of these galaxies might be M31 analogues, but much deeper
observations are needed to determine this.

       M33 is not an L$_*$ galaxy, with a luminosity of only about
20\% that of the Milky Way; in fact, at the mean distance of galaxies
in the survey, M33 would not have satisfied the optical 
brightness selection criterion to be included in the BIMA SONG sample.
However, we may still ask whether it would have
been detected at the mean distance of the survey.  The peak CO surface
density is about 1.5  M$_{\sun}$ pc$^{-2}$ \citep{Engargiola02} when
averaged in radial bins 500 pc in width.  Although this surface density
is well below the survey sensitivity, would it be possible to detect some
of the brightest of the individual GMCs?  In M33, there is no measurable
diffuse CO emission; the CO is concentrated in well defined GMCs.
The brightest of these GMCs has a mass of 6.6 $\times 10^5$ M$_{\sun}$
and has a mean diameter of 74 pc, so that the mean surface brightness
is 153 M$_{\sun}$ pc$^{-2}$.  At the mean distance of our sample, the
emission would be beam diluted to be 6.8 M$_{\sun}$ pc$^{-2}$ in a 350
pc beam, which is just below our sensitivity.  Because the emission is
in well-separated clouds, inclination effects would not be important
until the inclination became quite high.  However, M33 would be clearly
seen at the distance of some of the nearer galaxies.  At the distance of
4 Mpc, for example, the brightest cloud would have an observed surface
density of about 61 M$_{\sun}$ pc$^{-2}$, which is well above the SONG
detection threshold.  From the catalog of \citet{Engargiola02},
which lists 146 GMCs, 23 would be detected at the 3$\sigma$ level and
31 at 2$\sigma$ if M33 were at a distance of 4 Mpc and observed as part of
BIMA SONG.  Some of these might appear to be connected if their projected
separations were less than or about equal to the putative 120 pc beam
at 4 Mpc.

        The survey contains a close analogue to M33 both in optical
appearance and CO distribution: NGC 2403.  This galaxy is an Sc galaxy at a
distance of 4.2 Mpc, and is shown in the Hubble Atlas \citep{Sandage61}
on adjoining panels to emphasize their similarity.  The observed CO
distribution of NGC 2403 
is in fact consistent with originating in
individual GMCs like that in M33.  The total mass of M33, or 3 $\times$
10$^7$ M$_{\sun}$ \citep{Engargiola02}, is close to that of
NGC 2403, which we infer to
be 7 $\times$ 10$^7$ M$_{\sun}$ from \citet{Young95}.  At least
one SONG source has an inferred mass lower than M33: NGC 2976,
with 2 $\times$ 10$^7$ M$_{\sun}$.  Like NGC 2403, the CO emission
from NGC 2976 appears to be consistent with 
originating in individual GMCs.

        In summary, all of the Local Group spiral galaxies have CO
morphologies that are represented in the SONG survey.  However, the
CO luminosity of the Milky Way is somewhat below average, and for M31
and M33, they are well below average. 

\section{Conclusions}

We have presented the basic millimeter-wavelength data for the BIMA
Survey of Nearby Galaxies, an imaging survey of the 3 mm CO J = 1--0
emission within the centers and disks of 44 nearby spiral galaxies.
These are some of the main conclusions of the initial presentation
of the data.

1. Although the sample was not selected based on CO or infrared
brightness, we detected 41/44 or 93\% of the sources.  Together
with one nondetection that has known CO at radii outside the SONG
field of view, 42/44 = 95\% of the BIMA SONG galaxies have known CO
emission.  

2. We detected three of the sources (NGC 1068, NGC 3031, NGC 4579) 
in continuum emission at 112 GHz.  The continuum emission from
two of the sources (NGC 1068 and NGC 4579) appears marginally
resolved, while the emission from NGC 3031 (M81) is consistent
with a point source.

3. We showed that the fraction of total flux density recovered 
by the interferometer alone is a function of location within each map.
As predicted by imaging simulations, the fraction of total flux density 
recovered is also a function of the signal-to-noise ratio
of the interferometric data.  

4. While single-dish molecular gas surface densities show that most
molecular gas in nearby galaxies is confined to the central
$r \la$ 5 kpc, we showed that, on smaller scales, 
only 20/44 or 45\% of the SONG galaxies
have their peak molecular surface densities within the central
6\arcsec\ or $\sim$ 360 pc.   This suggests that it is in
fact quite common for molecular distributions {\em not} to
peak at the centers of spiral galaxies.

5. The three Local Group spiral galaxies have CO
morphologies that are represented in the SONG survey. 
The Milky Way would clearly have been detected at the average
distance of the survey, though it would have been below the
average brightness at that distance.  M31 would have been
marginally detectable at this distance, especially if its inclination angle
were higher than average.  The azimuthally-averaged molecular gas 
surface density in M33 is far below the detection threshold of the 
survey, but since the emission in M33 is concentrated in
well-defined clouds, we would have clearly detected emission from
these clouds if M33 were placed somewhat nearer than the average 
distance of the survey.

The data from this survey are publicly available at 
the Astronomy Digital Image Library (ADIL) at \\
http://adil.ncsa.uiuc.edu/document/02.TH.01
and at the NASA/IPAC Extragalactic Database (NED) at\\
http://nedwww.ipac.caltech.edu/level5/March02/SONG/SONG.html.  
The public data include 
the BIMA SONG integrated intensity images, channel maps, and
gaussian fit velocity fields.  Continuum images are included for the 
detected galaxies.  Associated files including the primary beam gain 
functions are also included, as are the single-dish
OTF channel maps and integrated intensity images.  A generic informational
file offers guidance to the released data.  The authors may be
contacted for further information.

\acknowledgements

We thank the BIMA Board for granting the observing time to complete
this Observatory project and the staffs at Hat Creek and the
National Radio Astronomy Observatory for their assistance
with the observations.  We thank John Lugten and Peter Teuben for
extensive discussions and assistance throughout this project.
This research was supported in part by NSF grants
AST 99-81308 (UC Berkeley) and AST 99-81289 (University of Maryland).
TTH was supported during some of the research period by NSF grant
AST 99-00789 (University of Arizona).

\appendix

\section{ Deconvolution Issues }

Interferometric imaging generally relies on nonlinear deconvolution
algorithms to help reconstruct the true source structure on the sky.
This is because, when the imaging algorithm grids the sampled visibility
data onto the $uv$ plane, it assigns all unsampled $uv$ cells the
unlikely value of zero.  The deconvolution algorithm works to find
a structure on the sky that is consistent with all the sampled
visibility data but that also provides a more plausible and robust
model of the unsampled visibility data.
Because the $uv$ coverage is inherently undersampled, there
may in principle be multiple source structures that are
consistent with the sampled $uv$ data, and the resultant
deconvolved image may not be a unique solution.  In practice,
different deconvolution techniques tend to give similar
results, which gives some reassurance as to the general structure
on the sky.  However, there are subtle differences among the
techniques that are especially apparent in the relatively
low signal-to-noise regime of millimeter-wavelength interferometry,
and we discuss these below.  For a introduction to the concept of 
the deconvolution as well as a detailed discussion of
the algorithms and how well they work, see the
excellent review by \citet{CBB99}.

The Maximum Entropy Method (MEM) works by fitting the sampled
$uv$ data to a model image that is designed to be positive and smooth where 
it is not otherwise constrained.  As \citet{CBB99} discuss,
one common pitfall in MEM is that the total flux density may
be seriously biased if the signal-to-noise ratio is low.
Using simulated
observations, \citet{HC99} showed that MEM
yields a reliable flux density when the signal-to-noise ratio is
above about 1000.  
However, with lower S/N ratios, MEM may introduce a 
positive bias in the model in the form of a large-scale plateau.  
This ``runaway'' flux problem gets progressively worse as the S/N 
ratio drops.  

\begin{figure}
\centerline{\includegraphics[height=6.5in,angle=-90]{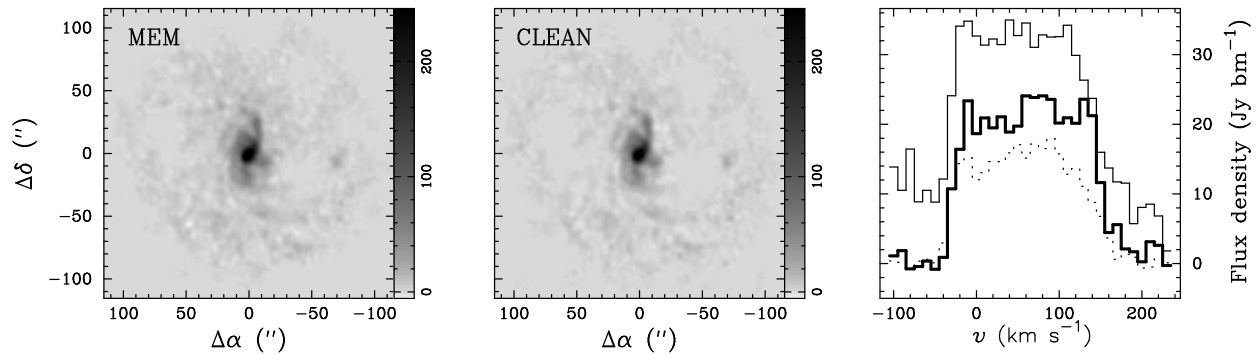} }
\caption{The ``runaway'' flux problem of MEM.
Shown are masked, BIMA-only maps of integrated intensity for NGC 6946, made
using MEM ({\it left}) and CLEAN ({\it middle}) deconvolutions.  While
the maps look generally very similar, the low-level flux density in
the MEM map has a positive bias.  This is shown in the spectra 
({\it right}), which show the total flux density measured in a
circle of 200\arcsec\ diameter for the single 
dish map (thick solid line), the CLEAN BIMA-only map
(dashed line), and the MEM BIMA-only map (thin solid line).   
The total flux in the MEM
map is over three times that in the CLEAN map; furthermore, it 
is (unphysically) over twice that of the corresponding single-dish flux.
}
\label{mosmem}
\end{figure}

In millimeter interferometry, the relatively low 
source strengths and relatively high random and systematic errors
conspire to ensure that we are almost always working in the
low signal-to-noise regime; for BIMA SONG, typical S/N ratios
for the channel maps were on the order of 10--40.
In BIMA SONG data, we saw the ``runaway'' flux feature commonly
with MEM, where there 
could be arbitrarily large flux density  added into the deconvolved
images, depending on what was specified as the convergence
criteria.  
An example of this is shown in Figure \ref{mosmem}, which shows
BIMA-only maps of integrated intensity for NGC 6946 made using
MEM and CLEAN, and spectra of the total flux measured in the
inner 200\arcsec-diameter region using the different
data cubes.  As seen in
the maps, the flux density in the small-scale components 
tends to be accurately represented, so that the final MEM and
CLEAN images look very similar to one another. 
But as the spectra show, the low-level flux rides systematically
higher with MEM than with CLEAN; in this example, the total flux in the
MEM map is over three times higher than that in the CLEAN map.
In fact, the total flux density in the BIMA-only MEM map is more than twice
the total flux density in the 12m-only map of the same region,
which is clearly an unphysical result.  In general, the deeper
one tries to deconvolve using MEM, the worse the ``runaway'' flux
problem gets.
The MEM algorithm performs somewhat better on a linear combination of 
the BIMA and 12m data, as with the \citet{Stan99} method described in 
Appendix B.
In principle, one could fine-tune the
convergence criteria for each channel of each map to try to
get an accurate flux determination.  However, there are pitfalls
to doing so:  first, if one cuts the deconvolution off too soon,
then residual sidelobes might remain in the map; second, specifying
the convergence criteria depends on {\it a priori} knowledge of
the source structure and flux density.  In practice it is difficult
to fine-tune MEM to work well in the low signal-to-noise regime.

The CLEAN algorithm, first introduced by \citet{Hogbom}, is an
iterative procedure that builds a model of the source distribution
using an ensemble of point sources.  \citet{CBB99} discuss the
many variations on the original algorithm contributed by
different authors; in BIMA SONG tests, we used the implementation
by \citet{SDI84}.
Since CLEAN does not suffer from the ``runaway'' flux
problem that MEM is subject to, 
we concluded that it was a far more reliable algorithm to use
for accurate flux density  representation for BIMA SONG maps; 
therefore, the maps in this catalog were produced using CLEAN.

While CLEAN does not introduce flux density  errors into the map, it also
does not produce an ideal deconvolution.  As 
\citet{HC99} showed, CLEAN does not do as well as 
MEM at recovering large-scale structure in the map. 
(These tests were done in the high signal-to-noise regime where 
MEM does not suffer from the ``runaway'' flux problem.)
This is because the conventional CLEAN source model 
is inherently built out of point sources, so that it is effectively 
biased against large-scale (and especially low-level) emission in a map,
whereas MEM inherently assumes a large-scale, smooth
distribution as a default model.

\begin{figure}
\centerline{\includegraphics[height=6.5in,angle=-90]{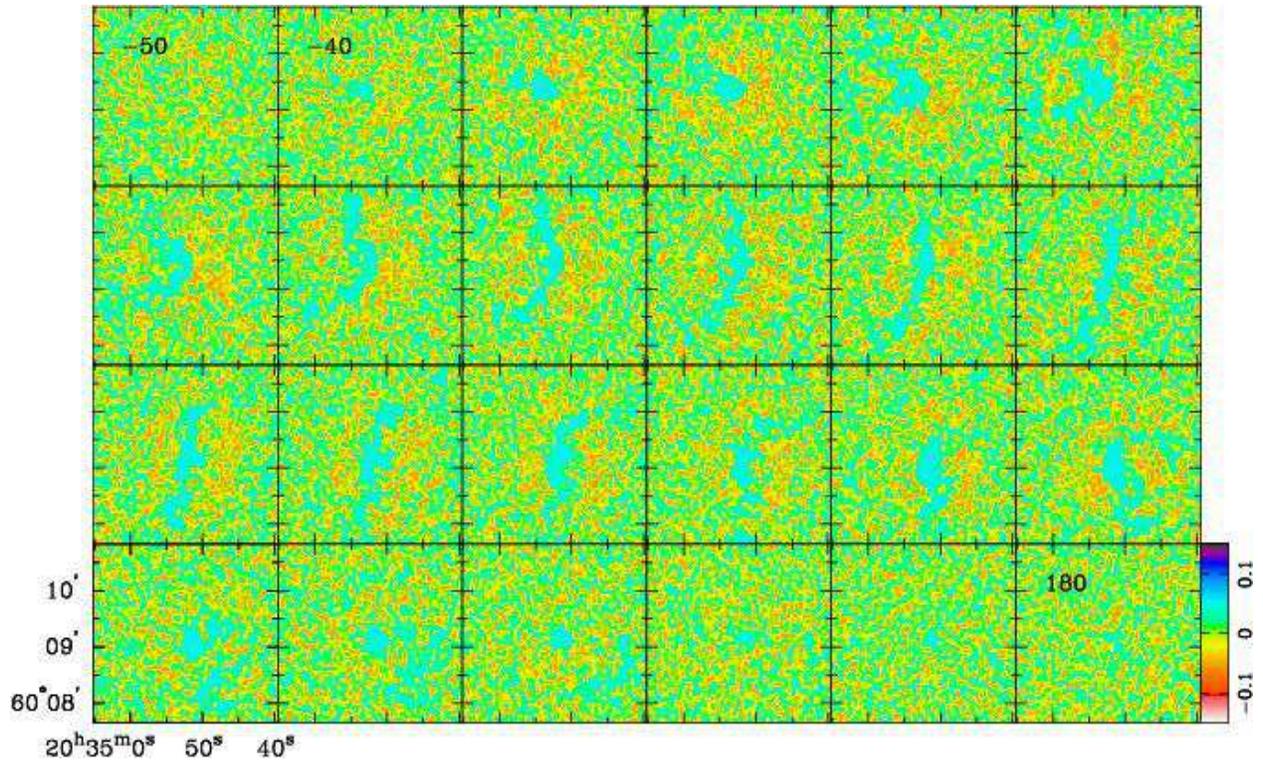} }
\caption{Residuals from the CLEAN deconvolution of the combined BIMA+12m
map of NGC 6946.  The residuals follow the general structure of the
channel maps shown in the Catalog figure for NGC 6946.  
Even for this very bright
source, the flux density in the residuals is about 8\% of the total 
flux density in
the final, ``restored'' map (which also includes the residuals).  
For a weaker source, the fraction of flux density in the residuals can
be much higher.
}
\label{reschan}
\end{figure}

A problem that is common to conventional CLEAN algorithms as well as to
MEM is that both algorithms tend to omit systematically
distributed residuals from the source models; the residuals,
even if at a level consistent with the 1 $\sigma$ noise specification,
take the general shape of the source distribution in each
channel map (Fig. \ref{reschan}).  This feature can introduce a bias for 
algorithms like 
self-calibration, which depend on the model to be an accurate 
representation of the source structure.  It can also lead to
an inaccurate flux density  determination in the final map:  the final
(``restored'') map is usually made by convolving the source model
(i.e. the output from the deconvolution algorithm) with an
idealized, gaussian beam, and then adding the derived residuals
to the result.  The smoothed model map therefore has units of 
flux density per ``clean beam'', whereas the residual map has units
of flux density per ``dirty beam''.  If one is at all fortunate,
then the dirty beam has an area that is not too dissimilar to that
of the clean beam, so that the convolved model and the residuals are
on the same flux scale.  But if there is a significant contribution
to the total flux density  that remains in the residual map, as may occur
when there is a broad, low-level source, and if the clean beam and
the dirty beam have slightly different areas, then the total flux density 
in the restored map might again be inaccurately represented.
The best one can do to alleviate this problem is to clean as deeply
as practical, so that as much of the flux density is represented
in the source model as is possible.

To summarize, (1) the CLEAN algorithm is probably a better choice
than MEM for imaging sources with S/N ratios $\la$ 1000, which
covers most non-masing sources at millimeter wavelengths with
current telescopes.
MEM may produce images with the correct peak flux density, but it may also
produce images with unphysically high total flux density  because of the
``runaway'' flux problem.  (2) It is extremely advantageous to get as 
much of the source flux density  into the deconvolution {\it model} as 
is possible, by cleaning as deeply as is practical.  That way, even if 
the model and the residuals are
on different flux density  scales, it is only the noise measurement that
is ill-defined.  The signal-to-noise estimates may then be affected,
but the source flux density  would be properly represented.

\section{ Single Dish and Interferometric Data Combination Techniques }

While the idea of incorporating single dish data into interferometric
data is not new (cf Emerson 1974 for the first 
combined image -- H I in M33, and Vogel et al.~1984 for the first
combined mm-$\lambda$ image -- HCO$^+$ in Orion), the practice
is far from routine.
The 24 combined BIMA$+$12 m maps presented in this paper comprise by
far the largest collection of single dish $+$ interferometric maps
published to date.  However, millimeter astronomers recognize the
importance of incorporating single dish data for making accurate
maps of large ($\ga$ 20--30\arcsec\ at 115 GHz) structures:  for the 
international Atacama Large Millimeter Array (ALMA) that is currently 
under development, the twin capabilities of mosaicing many fields
on the sky and of incorporating single-dish data into interferometric
maps together drive some of the most ambitious technical specifications
for the antenna design. 

\citet{Holdaway99} has presented an excellent, 
detailed discussion of techniques of combining mosaiced data with 
total power data, most of which have been implemented and tested
on high S/N data taken at centimeter wavelengths.  
Indeed, in noise-free simulations by \citet{Helfer2002}, we recovered 
all the flux density  in a resolved-out source
by simply adding a discrete total power point measured at ($u,v$)=(0,0)
and relying on the deconvolution to do the proper extrapolation
between the total power point and the interferometric data.
With real millimeter data, problems with pointing offsets and drifts, 
relative calibration between the telescopes, phase noise and
atmospheric decorrelation in the interferometer data,
and deformation of the telescope dish and therefore the
primary beam, combine to place severe constraints on the
deconvolution.  With the added difficulty that the signal-to-noise 
ratios are typically much lower, data combination tends to be much 
more challenging in the
millimeter regime than at centimeter wavelengths.     

For BIMA SONG, we tried a variety
of data combination techniques to determine what
worked best empirically for our data.  For many of our tests,
it was possible to fine-tune a single spectral channel
to achieve seemingly good pointing or calibration agreement,
but there would be a fairly random scatter of such solutions
across the 20-40 spectral channels that together made up a
data cube.  We therefore tried to achieve a solution that gave
the best results for the data cube as a whole.
 
\subsection{ Addition of Short- and Zero-Spacing Data in the $uv$ Plane }

In this technique, first demonstrated on 3 mm spectral line data by 
\citet{Vogel84}, the single-dish data are sampled at spatial
frequencies 0 $\le$ $u,v$ $\la$ $D_{\lambda}$, where $D_{\lambda}$
is the diameter of the single dish telescope.  In brief,
an idealized single-dish primary beam response function
is deconvolved from the single-dish OTF maps, to form a
model of what the source looks like on the sky.  For each
of the interferometer pointing centers, a model field is then
generated by multiplying the sky model by the idealized interferometer 
primary beam at that position.  The model field is then sampled at 
chosen $uv$ points.

In practice, this technique can be fine-tuned to work very well,
if one has enough patience. 
Unfortunately, the exact results tend to depend on
the details of the parameters chosen, from the method for
the single-dish deconvolution to the distribution, number and 
weighting of the $uv$ points specified (which can significantly
alter the synthesized beam from the interferometer-only case).   
As a check, one can
model visibilities just in the overlap region of the $uv$ plane
that both the single dish and the interferometer sample and
compare the calibration and structures in the resulting
maps of the overlap visibilities \citep[e.g.][]{WW94}; however, 
in relatively low S/N observations, this comparison of the overlap region
is usually unsatisfactory.  In the end, one can gauge the
success of the data combination by comparing the final combined
map, convolved to the resolution of the single-dish telescope,
to the single-dish data alone.   If by inspection the agreement
is not good, one can go back, adjust the calibration or try
a different (deeper or more shallow) single-dish deconvolution,
and try again.  In summary, 
while adding single-dish data in the $uv$ plane can be
very successful given enough attention to the combination,
it is typically time-consuming and hard to automate.

\subsection{ Addition of Single-Dish Data in the Image Plane}

There are both linear and nonlinear methods of adding the single-dish
data to an interferometric map in the image plane.  In the linear
method (implemented as IMMERGE in MIRIAD), the interferometric data are 
deconvolved separately.  The separate single-dish and interferometer
images are each transformed and then merged in the Fourier plane,
using a tapering function in the overlap region to specify 
the weights of each data set.  
One can check the goodness of
fit in the overlap region by comparing the $uv$ data derived from each
of the two input images.  However, with the S/N ratios
of 10--40 per spectral channel that were typical for BIMA SONG data,
the calibration in the overlap region depended rather sensitively 
on the exact region chosen for the overlap.
One also needs to be cautious when a large amount of flux density  is
resolved out by the interferometer; in this case, the interferometer-only
deconvolution may run into difficulties.  In practice, \citet{Wong2000}
used this technique successfully for CO-rich BIMA SONG galaxies.

\citet{SSB96} presented a nonlinear joint deconvolution algorithm
(implemented as MOSMEM in MIRIAD)
that Sault later modified to incorporate single dish data.  In
this method, a joint maximum entropy deconvolution is performed on a
linear mosaic of the ``dirty'' (undeconvolved) interferometer fields
along with the single-dish data; the algorithm maximizes the entropy,
subject to two separate $\chi^2$ constraints from the two data sets. 
This technique has been applied very successfully for
centimeter-wavelength data and for high S/N millimeter-wavelength
data \citep[e.g.][]{Welch2000}.
However, given the low S/N ``runaway'' flux problem associated with the
maximum entropy deconvolution and described in Appendix A,
this method did not turn out to be a successful approach for BIMA
SONG data.

Instead, for BIMA SONG, we adopted a linear combination technique
described in \citet{Stan99}.  In this method, we created a new
``dirty'' map by a linear combination of the interferometer map
prior to deconvolving and the single dish map, where the single
dish map had been weighted by an idealized form of the interferometer's
primary gain of the mosaiced pointings. 
We scaled the single dish map by the ratio of the area of the
interferometer beam to the area of the single dish beam, and created
a new ``dirty'' beam by the same linear combination of the
interferometer synthesized beam and the single dish beam (assumed
to be a truncated gaussian).  We then deconvolved the new combined
``dirty'' map using a Steer-Dewdney-Ito CLEAN algorithm. 
This method is easy to implement and automate, with relatively
few parameters to specify.

\clearpage

\begin{deluxetable}{lcccccccccccc}  
\tablecolumns{11}  
\tabletypesize{\scriptsize}
\rotate
\tablewidth{0pc}  
\tablecaption{ BIMA SONG Galaxy Sample }
\tablehead{  
\colhead{Source} &
\colhead{$\alpha$(J2000)\tablenotemark{a}} &   
\colhead{$\delta$(J2000)\tablenotemark{a}} &
\colhead{$v_{LSR}$\tablenotemark{a}} &
\colhead{$i\tablenotemark{b}$} &
\colhead{PA\tablenotemark{b}} &
\colhead{RC3 Type} &
\colhead{Nuclear\tablenotemark{c}} &
\colhead{$d$} &
\colhead{Distance} &
\colhead{$D_{25}$} &
\colhead{$B_T$} &
\colhead{Arm Class\tablenotemark{d}}\\

\colhead{} &
\colhead{($h$ $m$ $s$)} &
\colhead{(\arcdeg\ \arcmin\ \arcsec)} &
\colhead{(km~s$^{-1}$)} &
\colhead{(\arcdeg)} &
\colhead{(\arcdeg)} &
\colhead{} &
\colhead{Classification} &
\colhead{(Mpc)} &
\colhead{Reference} &
\colhead{(\arcmin)} &
\colhead{mag} &
\colhead{}
} 
\startdata  
NGC 0628 & 01 36 41.7 & $+$15 46 59 & 657    & 24\tablenotemark{e} & 25\tablenotemark{e} & SA(s)c & \nodata & 7.3 & 1      & 10.5 &  9.95 & 9\\
NGC 0925 & 02 27 16.9 & $+$33 34 45 & 553    & 56 & 102 & SAB(s)d & HII & 9.29 & 2 & 10.5 & 10.69 & 1\\
NGC 1068 & 02 42 40.7 & $-$00 00 48 & 1136   & 33\tablenotemark{e} & 13\tablenotemark{e}  & (R)SA(rs)b & Sy1.8 &14.4 & 1  & 7.1 & 9.61 & 3 \\
IC 342  & 03 46 49.7 & $+$68 05 45 & 34     & 31\tablenotemark{f} & 37\tablenotemark{f} &  SAB(rs)cd & HII  & 3.9 & 1 &         21.4  & 9.10  & 9\\
NGC 2403 & 07 36 54.5 & $+$65 35 58 & 131    & 56 & 127 & SAB(s)cd  & HII  & 4.2 & 1 &         21.9 &  8.93 & 3\\
NGC 2841 & 09 22 02.7 & $+$50 58 36 & 638    & 64 & 147 &  SA(r)b: & L2  & 14.1 & 3 &         8.1 & 10.09 &  3\\ 
NGC 2903 & 09 32 10.1 & $+$21 30 02 & 556    & 61\tablenotemark{e} & 17\tablenotemark{e} & SAB(rs)bc & HII & 6.3 & 1 & 12.6 &  9.68 & 7  \\ 
NGC 2976 & 09 47 17.1 & $+$67 54 51 &  3     & 63 & 143 & SAc-pec &  HII  & 2.1 & 1 &         5.9 & 10.82 & 3\\ 
NGC 3031 & 09 55 33.2 & $+$69 03 55 & $-$34  & 58 & 157 &  SA(s)ab & Sy1.5 & 3.63 & 4 &       26.9 &  7.89 & 12\\ 
NGC 3184 & 10 18 16.7 & $+$41 25 27 & 592    & 21 & 135 &  SAB(rs)cd & HII & 8.7  & 1 &         7.4 & 10.36 & 9\\
NGC 3344 & 10 43 31.1 & $+$24 55 21 & 586    & 24\tablenotemark{g} & 0\tablenotemark{g} & (R)SAB(r)bc & HII & 6.1 & 1 &         7.1 & 10.45 & 9\\ 
NGC 3351 & 10 43 58.0 & $+$11 42 14 & 778    & 40\tablenotemark{e} & 13\tablenotemark{e}  &  SB(r)b & HII & 10.1 & 5 &        7.4 & 10.53 & 6\\ 
NGC 3368 & 10 46 45.6 & $+$11 49 18 & 897    & 46\tablenotemark{g} & 5\tablenotemark{g} &  SAB(rs)ab & L2 & 11.2 & 6 &         7.6 & 10.11 & 8\\ 
NGC 3521 & 11 05 49.3 & $-$00 02 02 & 805    & 58\tablenotemark{e} & 164\tablenotemark{e} &  SAB(rs)bc & HII/L2:: & 7.2 & 1 &   11.0 & 9.83 & 3\\ 
NGC 3627 & 11 20 15.1 & $+$12 59 22 & 727    & 63\tablenotemark{e} & 176\tablenotemark{e} &  SAB(s)b & T2/Sy2 & 11.1 & 7 & 9.1 & 9.65 &7\\ 
NGC 3726 & 11 33 20.7 & $+$47 01 41 & 849    & 46 & 10 &  SAB(r)c & HII & 17.0 & 1 &     6.2 & 10.91 & 5\\ 
NGC 3938 & 11 52 49.6 & $+$44 07 14 & 809    & 24\tablenotemark{g} & 0\tablenotemark {g}    &  SA(s)c &  HII::  & 17.0 & 1 &         5.4 & 10.90 &  9\\ 
NGC 3953 & 11 53 49.0 & $+$52 19 37 & 1054   & 60 & 13 &  SB(r)bc & T2 & 17.0 & 1 & 6.9 & 10.84 & 9\\ 
NGC 3992 & 11 57 36.1 & $+$53 22 29 & 1048   & 52 & 68 &  SB(rs)bc & T2: & 17.0 & 1 &         7.6 & 10.60 &  9\\ 
NGC 4051 & 12 03 09.6 & $+$44 31 53 & 725    & 41 & 135 &  SAB(rs)bc & Sy1.2 & 17.0 & 1 &     5.2 & 10.83 & 5\\
NGC 4258 & 12 18 57.5 & $+$47 18 14 & 448    & 65\tablenotemark{e} & 176\tablenotemark{e} &  SAB(s)bc & Sy1.9 & 8.1 & 8 & 18.6 &  9.10 &\nodata \\ 
NGC 4303 & 12 21 54.9 & $+$04 28 25 & 1566   & 27\tablenotemark{g} & 0\tablenotemark{g}  &  SAB(rs)bc & HII & 15.2 & 1 &  6.5 & 10.18 &  9\\ 
NGC 4321 & 12 22 54.8 & $+$15 49 20 & 1571   & 32\tablenotemark{e} & 154\tablenotemark{e} &  SAB(s)bc & T2 & 16.1 & 9 & 7.4 & 10.05 & 12\\
NGC 4414 & 12 26 27.2 & $+$31 13 24 & 716    & 55\tablenotemark{e}& 159\tablenotemark{e} &  SA(rs)c? & T2:  & 19.1 & 10 &  3.6 & 10.96 & 3\\
NGC 4450 & 12 28 29.3 & $+$17 05 07 & 1954   & 41 & 175 &  SA(s)ab  & L1.9 & 16.8  &1 &         5.2 & 10.90 & 12\\ 
NGC 4490 & 12 30 36.9 & $+$41 38 23 & 565    & 60 & 125 &  SB(s)d pec & HII & 7.8 & 1 &         6.3 & 10.22 & 1\\ 
NGC 4535 & 12 34 20.3 & $+$08 11 54 & 1961   & 45 & 28\tablenotemark{h} &  SAB(s)c  & HII & 16.0 & 11 &         7.1 & 10.59 & 9\\ 
NGC 4548 & 12 35 26.4 & $+$14 29 49 & 486    & 37 & 150 &  SB(rs)b & L2 & 15.9 & 12 &      5.4 & 10.96 &5\\ 
NGC 4559 & 12 35 57.8 & $+$27 57 36 & 815    & 66 & 150 &  SAB(rs)cd & HII & 9.7 & 1 &         10.7 & 10.46 & \nodata  \\ 
NGC 4569 & 12 36 49.80 & $+$13 09 46 & $-$235 & 62 & 23 &  SAB(rs)ab & T2 &16.8 & 1 & 9.5 & 10.26 & 5\\
NGC 4579 & 12 37 43.53 & $+$11 49 05 & 1519   & 37 & 95 &  SAB(rs)b & Sy1.9/L1.9 &16.8  & 1 & 5.9 & 10.48 & 9\\ 
NGC 4699 & 12 49 02.33 & $-$08 39 55 & 1427   & 47 & 45 &  SAB(rs)b & \nodata & 25.7 & 1 & 3.8 & 10.41 & 3\\
NGC 4725 & 12 50 26.60 & $+$25 30 06 & 1206   & 45 & 35 &  SAB(r)ab pec & Sy2: & 12.6 & 13 &  10.7 & 10.11 & 6\\ 
NGC 4736 & 12 50 53.06 & $+$41 07 14 & 308    & 35\tablenotemark{e} & 100\tablenotemark{e} & (R)SA(r)ab & L2 & 4.3 & 1 &   11.2 &  8.99 & 3\\ 
NGC 4826 & 12 56 44.24 & $+$21 41 05 & 408    & 54\tablenotemark{e} & 111\tablenotemark{e} & (R)SA(rs)ab & T2 & 4.1 & 1 &     10.0 &  9.36 & 6\\ 
NGC 5005 & 13 10 56.23 & $+$37 03 33 & 946    & 61 & 65 &  SAB(rs)bc & L1.9 & 21.3  &1 & 5.8 & 10.61 & 3\\ 
NGC 5033 & 13 13 27.53 & $+$36 35 38 & 875    & 62 & 170 &  SA(s)c & Sy1.5  & 18.7 & 1 & 10.7 & 10.75 &  9\\ 
NGC 5055 & 13 15 49.25 & $+$42 01 49 & 504    & 55 & 105 &  SA(rs)bc & T2 & 7.2 & 1 & 12.6 &  9.31 & 3\\ 
NGC 5194 & 13 29 52.35 & $+$47 11 54 & 463    & 15\tablenotemark{e} & 0\tablenotemark{e}  &  SA(s)bc pec & Sy2 & 7.7 & 1 &  11.2 &  8.96 & 12\\ 
NGC 5248 & 13 37 32.06 & $+$08 53 07 & 1153   & 43 & 110 &  SAB(rs)bc & HII & 22.7 & 1 &  6.2 & 10.97 & 12\\
NGC 5247 & 13 38 02.62 & $-$17 53 02 & 1357   & 29 & 20 &  SA(s)bc  & \nodata & 22.2  & 1 & 5.6 & 10.5  & 9\\
NGC 5457 & 14 03 12.48 & $+$54 20 55 & 241    & 27\tablenotemark{i} & 40\tablenotemark{i} &  SAB(rs)cd & \nodata  & 7.4 & 14 &         28.8 &  8.31  & 9\\ 
NGC 6946 & 20 34 52.33 & $+$60 09 14 &  48    & 54\tablenotemark{e} & 65\tablenotemark{e} &  SAB(rs)cd & HII & 5.5 & 1 &      11.5 &  9.61 & 9\\ 
NGC 7331 & 22 37 04.09 & $+$34 24 56 & 821    & 62\tablenotemark{e} & 172\tablenotemark{e}&  SA(s)b & T2  & 15.1 & 15 & 10.5 & 10.35  & 3\\ 
\enddata
\tablenotetext{a}{Adopted tracking center and LSR velocity of the observations.}
\tablenotetext{b}{Inclination ($i$) and position angle (PA) from RC3, 
except where noted.}
\tablenotetext{c}{Spectral classification from \citet{HFS97}. HII = H II 
nucleus, Sy = Seyfert nucleus, L = LINER, and T = transition object,
which has [O I] strengths intermediate between those of H II nuclei and
LINERs.  One (:) or two (::) colons following the classification signifies
uncertain and highly uncertain classifications.  
}  
\tablenotetext{d}{Arm class (AC) from \citet{EE87}. AC 1--4 are flocculents,
and AC 5--12 are grand design spirals.}
\tablenotetext{e}{From Paper I.}
\tablenotetext{f}{From kinematic fitting, \citet{CTH00}.}
\tablenotetext{g}{From optical photometry, \citet{Frei96}.}
\tablenotetext{h}{From near-infrared photometry, \citet{MH01}.}
\tablenotetext{i}{From kinematic fitting, \citet{Braun97}.}
\tablerefs{(1) \citet{Tully88};(2) \cite{Silb96}; (3) \citet{Macri01};
(4) \citet{Freed94}; (5) \citet{Graham97}; (6) \citet{TFS99};
(7) \citet{Saha99}; (8) \citet{Maoz99}; (9) \citet{Ferrarese96}; 
(10) \citet{Turner98}; (11) \citet{Macri99};
(12) \citet{Graham99}; (13) \citet{Gibson99};
(14) \citet{Kelson96}; (15) \citet{Hughes98}
}
\end {deluxetable}
\clearpage

\begin{deluxetable}{lccccccccc}  
\tablecolumns{10}  
\tabletypesize{\tiny}
\tablewidth{0pc}  
\tablecaption{Observational parameters}
\tablehead{  
\colhead{Source} &
\colhead{N$_{fields}$\tablenotemark{a}} &   
\colhead{$\theta_{maj}\times\theta_{min}$\tablenotemark{b}} &   
\colhead{$D_{maj}\times D_{min}$\tablenotemark{c}} &
\colhead{$bpa$\tablenotemark{d}} &
\colhead{$\Delta v$\tablenotemark{e}} &
\colhead{$\sigma_{10}$\tablenotemark{f}} &
\colhead{Cont.~flux\tablenotemark{g}} &
\colhead{Contour\tablenotemark{h}} &
\colhead{$\sigma_{12m}$\tablenotemark{i}}\\

\colhead{} &
\colhead{} &
\colhead{(\arcsec $\times$ \arcsec)} &
\colhead{(pc $\times$ pc)} &
\colhead{(\arcdeg)} &
\colhead{(km~s$^{-1}$)} &
\colhead{(mJy~bm$^{-1}$)} &
\colhead{(Jy~bm$^{-1}$)} &
\colhead{(Jy~bm$^{-1}$~km~s$^{-1}$)} &
\colhead{(Jy~bm$^{-1}$)}
} 
\startdata  
NGC 0628 & 13 & \phn 7.2 $\times$ 5.3 & \phn 250 $\times$  190 & \phn\phs 8 & 5 & \phn 51 & $<$0.012 & 1.5 & 0.69 \\
NGC 0925 & 7 & \phn 6.1 $\times$ 5.6 & \phn 280 $\times$  250 & \phn$-$4 & 10 & \phn 65 & $<$0.013 & 2.5 & \nodata \\
NGC 1068 & 13 & \phn 8.9 $\times$ 5.6 & \phn 620 $\times$  390 & \phs 16 & 10 & \phn 85 & 0.041 $\pm$ 0.007 & 6.0 & 0.56 \\
IC 342 & 7 & \phn 5.6 $\times$ 5.1 & \phn 110 $\times$ \phn 97 & \phs 34 & 10 &  129 & $<$0.027 & 6.0 & 1.47 \\
NGC 2403 & 7 & \phn 6.3 $\times$ 5.8 & \phn 130 $\times$  120 & \phs 22 & 10 & \phn 49 & $<$0.009 & 2.0 & \nodata \\
NGC 2841 & 7 & \phn 6.1 $\times$ 5.1 & \phn 420 $\times$  350 & \phs 25 & 20 & \phn 66 & $<$0.013 & 2.5 & \nodata \\
NGC 2903 & 7 & \phn 6.8 $\times$ 5.5 & \phn 210 $\times$  170 & \phn\phs 9 & 10 & \phn 68 & $<$0.016 & 5.0 & 0.38 \\
NGC 2976 & 1 & \phn 7.8 $\times$ 6.9 & \phn\phn 79 $\times$ \phn 70 & \phs 59 & 10 & \phn 41 & $<$0.010 & 1.5 & \nodata \\
NGC 3031 & 7 & \phn 5.9 $\times$ 5.1 & \phn 100 $\times$ \phn 90 & $-$14 & 20 & \phn 58 & 0.161 $\pm$ 0.006 & 3.0 & \nodata \\
NGC 3184 & 7 & \phn 5.9 $\times$ 5.4 & \phn 250 $\times$  230 & \phs 12 & 10 & \phn 50 & $<$0.010 & 1.5 & \nodata \\
NGC 3344 & 7 & \phn 6.4 $\times$ 5.3 & \phn 190 $\times$  160 & \phn\phs 9 & 10 & \phn 50 & $<$0.011 & 1.3 & \nodata \\
NGC 3351 & 7 & \phn 7.4 $\times$ 5.2 & \phn 360 $\times$  250 & \phn\phs 7 & 10 & \phn 64 & $<$0.014 & 3.0 & 0.33 \\
NGC 3368 & 7 & \phn 7.7 $\times$ 5.7 & \phn 420 $\times$  310 & \phn\phs 5 & 10 & \phn 69 & $<$0.013 & 2.0 & \nodata \\
NGC 3521 & 7 & \phn 8.8 $\times$ 5.7 & \phn 310 $\times$  200 & \phn\phs 1 & 10 & \phn 78 & $<$0.020 & 5.0 & 0.56 \\
NGC 3627 & 14 & \phn 7.3 $\times$ 5.8 & \phn 390 $\times$  310 & \phs 80 & 10 & \phn 41 & $<$0.009 & 5.0 & 0.43 \\
NGC 3726 & 1 & \phn 6.0 $\times$ 5.0 & \phn 500 $\times$  420 & \phs 11 & 10 & \phn 32 & $<$0.008 & 1.5 & \nodata \\
NGC 3938 & 7 & \phn 5.9 $\times$ 5.4 & \phn 480 $\times$  440 & $-$33 & 10 & \phn 59 & $<$0.014 & 2.5 & 0.31 \\
NGC 3953 & 7 & \phn 6.9 $\times$ 5.8 & \phn 570 $\times$  480 & \phn\phs 2 & 10 & \phn 72 & $<$0.015 & 2.0 & \nodata \\
NGC 3992 & 7 & \phn 6.6 $\times$ 5.3 & \phn 550 $\times$  440 & \phn\phs 1 & 10 & \phn 60 & $<$0.013 & 1.5 & \nodata \\
NGC 4051 & 1 & \phn 7.0 $\times$ 4.9 & \phn 580 $\times$  410 & \phn$-$2 & 10 & \phn 23 & $<$0.006 & 1.5 & \nodata \\
NGC 4258 & 7 & \phn 6.1 $\times$ 5.4 & \phn 240 $\times$  210 & \phn\phs 4 & 10 & \phn 60 & $<$0.013 & 4.0 & 0.26 \\
NGC 4303 & 7 & \phn 7.3 $\times$ 5.5 & \phn 540 $\times$  400 & \phn$-$4 & 10 & \phn 47 & $<$0.011 & 4.0 & 0.66 \\
NGC 4321 & 7 & \phn 7.2 $\times$ 4.9 & \phn 560 $\times$  380 & \phn\phs 9 & 10 & \phn 50 & $<$0.014 & 3.0 & 0.94 \\
NGC 4414 & 1 & \phn 6.4 $\times$ 5.0 & \phn 590 $\times$  460 & \phn\phs 6 & 10 & \phn 29 & $<$0.008 & 3.0 & 0.36 \\
NGC 4450 & 1 & \phn 6.6 $\times$ 5.1 & \phn 540 $\times$  410 & \phn\phs 5 & 10 & \phn 43 & $<$0.013 & 1.5 & \nodata \\
NGC 4490 & 1 & \phn 6.0 $\times$ 5.7 & \phn 230 $\times$  210 & \phs 58 & 10 & \phn 26 & $<$0.007 & 0.8 & \nodata \\
NGC 4535 & 7 & \phn 7.3 $\times$ 5.7 & \phn 570 $\times$  440 & \phn\phs 2 & 10 & \phn 63 & $<$0.015 & 2.0 & \nodata \\
NGC 4548 & 1 & \phn 7.1 $\times$ 6.2 & \phn 550 $\times$  480 & $-$23 & 10 & \phn 43 & $<$0.011 & 2.0 & \nodata \\
NGC 4559 & 7 & \phn 5.9 $\times$ 5.6 & \phn 280 $\times$  260 & $-$16 & 10 & \phn 55 & $<$0.013 & 2.0 & \nodata \\
NGC 4569 & 7 & \phn 6.6 $\times$ 5.5 & \phn 540 $\times$  450 & \phn$-$2 & 10 & \phn 54 & $<$0.012 & 3.0 & 0.24 \\
NGC 4579 & 7 & \phn 8.1 $\times$ 7.1 & \phn 660 $\times$  580 & $-$11 & 10 & \phn 68 & 0.028 $\pm$ 0.006 & 3.5 & \nodata \\
NGC 4699 & 1 & 10.3 $\times$ 6.8 &  1300 $\times$  850 & \phs 11 & 10 & \phn 86 & $<$0.019 & 3.0 & \nodata \\
NGC 4725 & 7 & \phn 7.0 $\times$ 5.4 & \phn 430 $\times$  330 & \phn\phs 9 & 20 & \phn 86 & $<$0.017 & 3.0 & \nodata \\
NGC 4736 & 7 & \phn 6.9 $\times$ 5.0 & \phn 140 $\times$  100 & \phs 63 & 10 & \phn 64 & $<$0.017 & 3.0 & 0.38 \\
NGC 4826 & 7 & \phn 7.5 $\times$ 5.2 & \phn 150 $\times$  100 & \phn\phs 3 & 10 & \phn 82 & $<$0.016 & 4.0 & 0.39 \\
NGC 5005 & 1 & \phn 6.2 $\times$ 6.0 & \phn 640 $\times$  620 & \phs 23 & 10 & \phn 32 & $<$0.009 & 1.5 & 0.16 \\
NGC 5033 & 7 & \phn 6.1 $\times$ 5.4 & \phn 550 $\times$  490 & \phs 86 & 20 & \phn 73 & $<$0.015 & 4.0 & 0.44 \\
NGC 5055 & 7 & \phn 5.8 $\times$ 5.5 & \phn 200 $\times$  190 & \phs 18 & 10 & \phn 55 & $<$0.011 & 5.0 & 0.45 \\
NGC 5194 & 26 & \phn 5.8 $\times$ 5.1 & \phn 220 $\times$  190 & \phn\phs 4 & 10 & \phn 61 & $<$0.012 & 5.0 & 0.55 \\
NGC 5248 & 1 & \phn 6.9 $\times$ 5.8 & \phn 760 $\times$  630 & \phn\phs 1 & 10 & \phn 38 & $<$0.011 & 1.0 & 0.35 \\
NGC 5247 & 1 & 12.2 $\times$ 4.7 &  1300 $\times$  510 & \phn\phs 2 & 10 & \phn 59 & $<$0.016 & 3.0 & 0.32 \\
NGC 5457 & 7 & \phn 5.7 $\times$ 5.4 & \phn 210 $\times$  190 & \phs 59 & 10 & \phn 45 & $<$0.010 & 3.0 & 0.59 \\
NGC 6946 & 7 & \phn 6.0 $\times$ 4.9 & \phn 160 $\times$  130 & \phs 14 & 10 & \phn 61 & $<$0.013 & 6.0 & 0.63 \\
NGC 7331 & 7 & \phn 6.1 $\times$ 4.9 & \phn 440 $\times$  360 & \phn\phs 5 & 10 & \phn 50 & $<$0.010 & 3.0 & 0.51 \\
\enddata
\tablenotetext{a}{Number of fields observed with the interferometer.}
\tablenotetext{b}{FWHM of major and minor axes of synthesized beam.}
\tablenotetext{c}{Linear dimensions of major and minor axes of synthesized
beam, given the distances in Table 1.}
\tablenotetext{d}{Beam position angle.}
\tablenotetext{e}{Velocity binning of the channel maps.}
\tablenotetext{f}{Rms noise level of the channel map, measured over a 
120\arcsec\ $\times$ 120\arcsec\ box 
in emission-free channels in the data cube.  Where 
the velocity binning $\Delta v$ was other than 10 km s$^{-1}$, the rms
was normalized to that of a 10 km s$^{-1}$ channel.
}
\tablenotetext{g}{Continuum flux levels.  Upper limits are 3 $\sigma$
measurements.  For the detections, formal uncertainties are listed;
the systematic error is 1 $\sigma$ = 15\%.}
\tablenotetext{h}{Starting contour for the maps of integrated intensity
in \S 5.}
\tablenotetext{i}{Rms noise level of single-dish OTF channel maps, per
10 km s$^{-1}$ channel.  Where present, 
this indicates that single-dish OTF maps from the NRAO 12 m telescope 
were incorporated into the BIMA SONG maps. }
\end{deluxetable}

\clearpage

\begin{deluxetable}{ccccccl}  
\tablecolumns{8}  
\tabletypesize{\footnotesize}
\tablewidth{0pc}  
\tablecaption{Position-Switched Data from NRAO 12m Telescope}  
\tablehead{  
\colhead{Source} &
\colhead{$\Delta\alpha$\tablenotemark{a}} &   
\colhead{$\Delta\delta$\tablenotemark{a}} &   
\colhead{$\sigma_R^*$\tablenotemark{b}} &
\colhead{$I_R^* \pm \sigma_I$\tablenotemark{c}} &
\colhead{$v_{min},v_{max}$\tablenotemark{d}} &
\colhead{Flux Fraction\tablenotemark{e}}\\

\colhead{} &
\colhead{(\arcsec)} &
\colhead{(\arcsec)} &
\colhead{(K)} &
\colhead{(K~km~s$^{-1}$)} &
\colhead{(km~s$^{-1}$)} &
\colhead{Recovered}
} 
\startdata  
NGC 0925 & 0 & 0 & 0.005 & 1.10 $\pm$ 0.15 &  495,640  & (0.07 $\pm$ 0.38) \\
 & 22 & 38 & 0.009 & $<$0.87  &  453,653  & \nodata \\
 & 44 & 0 & 0.013 & 0.76 $\pm$ 0.23 &  480,540 & \nodata \\
 & 22 & -38 & 0.005 & 1.25 $\pm$ 0.14 &  430,550  & (0.06 $\pm$ 0.30) \\
 & -22 & -38 & 0.006 & $<$0.57 & 453,653  & \nodata\\
 & -44 & 0 & 0.012 & $<$1.1 & 453,653  & \nodata\\
 & -22 & 38 & 0.007 & $<$0.69 & 453,653  & \nodata \\[10pt]

NGC 2403 & 0 & 0 & 0.018 & 2.16 $\pm$ 0.29 & 100,200  & 0.63 $\pm$ 0.16  \\
 & 22 & 38 & 0.023 & $<$1.5 & 31,231  & \nodata\\
 & 44 & 0 & 0.019 & 1.71 $\pm$ 0.25 &  145,210  & (0.19  $\pm$ 0.15) \\
 & 22 & -38 & 0.017 & 1.14 $\pm$ 0.27 &  100,200  & (0.55  $\pm$ 0.29) \\
 & -22 & -38 & 0.015 & 1.46 $\pm$ 0.25 & 90,200  & 0.56  $\pm$ 0.23  \\
 & -44 & 0 & 0.020 & $<$1.4 & 31,231  & \nodata\\
 & -22 & 38 & 0.020 & 1.95 $\pm$ 0.39 & 50,200  & (0.24  $\pm$ 0.18) \\[10pt]

NGC 2841 & 0 & 0 & 0.008 & $<$1.4 & 290,930  & \nodata\\
 & 22 & 38 & 0.010 & $<$1.7 & 290,930  & \nodata\\
 & 44 & 0 & 0.010 & $<$1.7 & 290,930  & \nodata\\
 & 22 & -38 & 0.011 & $<$1.9 & 290,930   & \nodata\\
 & -22 & -38 & 0.009 & $<$1.7 & 290,930  & \nodata\\
 & -44 & 0 & 0.011 & $<$1.8 & 290,930  & \nodata\\
 & -22 & 38 & 0.010 & $<$1.8 & 290,930  & \nodata\\[10pt]

NGC 2976 & 0 & 0 & 0.007 & 1.29 $\pm$ 0.15 &  -35,45  & (0.49 $\pm$ 0.42) \\[10pt]

NGC 3031 & 0 & 0 & 0.013 & $<$1.3 & -134,66  & \nodata\\
 & 22 & 38 & 0.021 & $<$2.0 & -134,66  & \nodata\\
 & 44 & 0 & 0.014 & $<$1.3 & -134,66  & \nodata\\
 & 22 & -38 & 0.014 & $<$1.4 & -134,66  & \nodata\\
 & -22 & -38 & 0.020 & $<$1.9 & -134,66  & \nodata\\
 & -44 & 0 & 0.015 & $<$1.5 & -134,66  & \nodata\\
 & -22 & 38 & 0.011 & 1.25 $\pm$ 0.28 & 70,180 &  \nodata\\[10pt]

NGC 3184 & 0 & 0 & 0.004 & 3.96 $\pm$ 0.09 & 550,660  & 0.54 $\pm$ 0.07 \\
 & 22 & 38 & 0.008 & 3.10 $\pm$ 0.17 &  535,620  & 0.28 $\pm$ 0.08  \\
 & 44 & 0 & 0.007 & 2.44 $\pm$ 0.14 &  560,635  & 0.30 $\pm$ 0.10  \\
 & 22 & -38 & 0.008 & 2.79 $\pm$ 0.17 &  550,645  & (0.17 $\pm$ 0.09) \\
 & -22 & -38 & 0.007 & 2.43 $\pm$ 0.14 &  570,650  & (0.16 $\pm$ 0.10) \\
 & -44 & 0 & 0.009 & 2.72 $\pm$ 0.22 &  510,630  & (0.05 $\pm$ 0.10) \\
 & -22 & 38 & 0.010 & 2.35 $\pm$ 0.24 & 500,600  & (0.07 $\pm$ 0.11)  \\[10pt]

NGC 3344 & 0 & 0 & 0.015 & 3.28 $\pm$ 0.45 & 475,650  & (0.01 $\pm$ 0.11) \\
 & 22 & 38 & 0.018 & 2.85 $\pm$ 0.53 &  490,660 &  \nodata\\
 & 44 & 0 & 0.016 & $<$1.1 &  500,600 &  \nodata\\
 & 22 & -38 & 0.014 & 1.90 $\pm$ 0.22 &  500,545  & (0.02 $\pm$ 0.10)\\
 & -22 & -38 & 0.019 & 3.34 $\pm$ 0.46 &  500,610  & (0.06 $\pm$ 0.08)\\
 & -44 & 0 & 0.010 & 2.25 $\pm$ 0.30 & 475,650  & (0.13 $\pm$ 0.15)\\
 & -22 & 38 & 0.013 & 2.28 $\pm$ 0.24 & 600,660  & 0.19 $\pm$ 0.09  \\[10pt]

NGC 3368 & 0 & 0 & 0.006 & 12.80 $\pm$ 0.32 &  670,1200  & 0.94 $\pm$ 0.06  \\
 & 22 & 38 & 0.009 & $<$0.90 & 797,997  & \nodata\\
 & 44 & 0 & 0.007 & 3.04 $\pm$ 0.29 & 800,1100  & 0.36 $\pm$ 0.17   \\
 & 22 & -38 & 0.007 & 3.32 $\pm$ 0.28 & 800,1100  & 0.73 $\pm$ 0.16  \\
 & -22 & -38 & 0.007 & 4.23 $\pm$ 0.30 & 750,1100  & 0.28 $\pm$ 0.13  \\
 & -44 & 0 & 0.008 & 3.54 $\pm$ 0.32 & 680,1000  & (0.27  $\pm$ 0.15) \\
 & -22 & 38 & 0.007 & 4.58 $\pm$ 0.27 & 680,1000  & 0.26 $\pm$ 0.11   \\[10pt]

NGC 3726 & 0 & 0 & 0.010 & 2.88 $\pm$ 0.33 & 760,960  & (0.43 $\pm$ 0.25) \\[10pt]

NGC 3953 & 0 & 0 & 0.010 & 4.57 $\pm$ 0.39 & 860,1170  & (0.10 $\pm$ 0.16) \\
 & 22 & 38 & 0.009 & 4.90 $\pm$ 0.32 & 1030,1250  & (0.06  $\pm$ 0.13)\\
 & 44 & 0 & 0.011 & 3.09 $\pm$ 0.41 & 950,1200  & (0.06  $\pm$ 0.22)\\
 & 22 & -38 & 0.009 & 3.99 $\pm$ 0.30 & 870,1100  & (0.05  $\pm$ 0.16)\\
 & -22 & -38 & 0.009 & 4.71 $\pm$ 0.25 & 850,1000  & (0.13 $\pm$ 0.11)\\
 & -44 & 0 & 0.010 & 5.21 $\pm$ 0.43 & 840,1170  & (0.05 $\pm$ 0.15)\\
 & -22 & 38 & 0.009 & 5.62 $\pm$ 0.33 & 1000,1240  & (0.06 $\pm$ 0.12)\\[10pt]

NGC 3992 & 0 & 0 & 0.008 & $<$1.0 & 800,1200   & \nodata\\
 & 22 & 38 & 0.008 & $<$1.1 & 800,1200   & \nodata\\
 & 44 & 0 & 0.007 & 1.54 $\pm$ 0.21 & 800,950  & (-0.22 $\pm$ 0.22)\\
 & 22 & -38 & 0.009 & $<$1.3 & 800,1200   & \nodata\\ 
 & -22 & -38 & 0.008 & $<$1.1 & 800,1200 & \nodata \\
 & -44 & 0 & 0.010 &  1.54 $\pm$ 0.45 & 800,1200 & \nodata\\
 & -22 & 38 & 0.009 & 2.06 $\pm$ 0.43 & 800,1200 & \nodata \\[10pt]

NGC 4051 & 0 & 0 & 0.012 & 8.68 $\pm$ 0.42 & 615,845  & 0.34 $\pm$ 0.06  \\[10pt]

NGC 4450 & 0 & 0 & 0.011 & 4.17 $\pm$ 0.46 & 1800,2150  & (0.18 $\pm$ 0.32) \\[10pt]

NGC 4490 & 0 & 0 & 0.004 & 1.63 $\pm$ 0.12 & 550,720  & (0.27 $\pm$ 0.34) \\[10pt]

NGC 4535 & 0 & 0 & 0.011 & 8.26 $\pm$ 0.45 & 1810,2120  & 0.65 $\pm$ 0.07  \\
 & 22 & 38 & 0.020 & 3.99 $\pm$ 0.56 & 1810,1960  & 0.29 $\pm$ 0.10  \\
 & 44 & 0 & 0.019 & 5.12 $\pm$ 0.64 & 1860,2090  & 0.31 $\pm$ 0.09  \\
 & 22 & -38 & 0.019 & 3.83 $\pm$ 0.49 & 1950,2080  & 0.21 $\pm$ 0.09  \\
 & -22 & -38 & 0.018 & 3.86 $\pm$ 0.49 & 1950,2095  & 0.31 $\pm$ 0.10  \\
 & -44 & 0 & 0.024 & 2.09 $\pm$ 0.61 & 1915,2040  & 0.41 $\pm$ 0.20  \\
 & -22 & 38 & 0.019 & 3.17 $\pm$ 0.50 & 1810,1945  & 0.31 $\pm$ 0.12   \\[10pt]

NGC 4548 & 0 & 0 & 0.005 & 2.77 $\pm$ 0.19 & 305,625  & (0.31 $\pm$ 0.42) \\[10pt]

NGC 4559 & 0 & 0 & 0.008 & 3.09 $\pm$ 0.26 & 715,910  & 0.31 $\pm$ 0.12  \\
 & 22 & 38 & 0.012 & $<$1.1 & 715,910   & \nodata\\
 & 44 & 0 & 0.011 & $<$1.0 & 715,910  & \nodata \\
 & 22 & -38 & 0.011 & 1.99 $\pm$ 0.28 & 690,825  & (0.26 $\pm$ 0.16) \\
 & -22 & -38 & 0.008 & $<$0.78 & 715,910  & \nodata\\
 & -44 & 0 & 0.011 & 1.97 $\pm$ 0.30 & 750,900  & (-0.04 $\pm$ 0.16) \\
 & -22 & 38 & 0.009 & 1.80 $\pm$ 0.23 & 790,910  & (0.22 $\pm$ 0.17) \\

NGC 4579 & 0 & 0 & 0.006 & 6.33 $\pm$ 0.26 & 1310,1705  & 0.32 $\pm$ 0.08  \\
 & 22 & 38 & 0.014 & 3.73 $\pm$ 0.62 & 1310,1705  & (0.17 $\pm$ 0.14)\\
 & 44 & 0 & 0.013 & 2.56 $\pm$ 0.36 & 1565,1705  & (0.13$\pm$ 0.12)\\
 & 22 & -38 & 0.014 & 2.09 $\pm$ 0.36 & 1525,1645 & \nodata\\
 & -22 & -38 & 0.015 & 5.62 $\pm$ 0.49 & 1320,1540  & (0.13 $\pm$ 0.07)\\
 & -44 & 0 & 0.019 & 3.89 $\pm$ 0.66 & 1320,1540 & \nodata\\
 & -22 & 38 & 0.020 & 3.71 $\pm$ 0.67 & 1320,1540 & \nodata\\[10pt]

NGC 4699 & 0 & 0 & 0.011 & $<$1.5 & 1250,1620   & \nodata\\[10pt]

NGC 4725 & 0 & 0 & 0.007 & 2.80 $\pm$ 0.32 & 1030,1400  & (0.60 $\pm$ 0.31)\\
 & 22 & 38 & 0.010 & $<$1.3 & 1030,1400   & \nodata\\
 & 44 & 0 & 0.018 & $<$2.3 & 1030,1400   & \nodata\\
 & 22 & -38 & 0.015 & $<$1.9 & 1030,1400   & \nodata\\
 & -22 & -38 & 0.015 & $<$1.9 & 1030,1400   & \nodata\\
 & -44 & 0 & 0.012 & $<$1.5 & 1030,1400   & \nodata\\
 & -22 & 38 & \nodata & \nodata & \nodata & \nodata \\
\enddata
\tablenotetext{a}{Location of spectra. Offsets are measured from tracking 
center listed in Table 1.}
\tablenotetext{b}{Measured rms noise per 2 MHz (5.2 km~s$^{-1}$) channel. }
\tablenotetext{c}{Integrated intensity $I_R^*$ = $\Sigma T_R^* \Delta v$, 
measured over the velocity limits given.  The formal uncertainty in the 
integrated intensity is also tabulated.  The systematic uncertainty has an 
assigned value of about 25\% times the integrated intensity; see \S 3.2.2.
For nondetections, a 3 $\sigma_I$ upper limit is given. }
\tablenotetext{d}{Velocity limits of integration.}
\tablenotetext{e}{Ratio of flux recovered in the BIMA spectra, smoothed to 
55\arcsec, relative to the 12m flux.  The formal uncertainty in the ratio
is listed.  Given the assigned systematic uncertainties of 15\% in
the BIMA data and 25\% in the 12 m data, the systematic uncertainty
in the ratio is 30\% times the ratio.  The ratios are enclosed in 
parentheses when they are detected at
$\le$ 2 $\sigma$ confidence level. See \S 8.3.}
\end{deluxetable}

\clearpage

\begin{deluxetable}{lccccc}  
\tablecolumns{6}
\tabletypesize{\scriptsize}
\tablewidth{0pc}  
\tablecaption{ Global Flux Densities, Luminosities, and Masses for OTF data}
\tablehead{  
\colhead{Source} &
\colhead{$\Delta\alpha$\tablenotemark{a}} &
\colhead{$\Delta\alpha/D_{25}$} &
\colhead{$S_{tot}$\tablenotemark{b}} &
\colhead{$L_{CO}$\tablenotemark{c}} &
\colhead{$M(H_2)$\tablenotemark{d}} \\
\colhead{} &
\colhead{(\arcmin)} &
\colhead{} &
\colhead{(Jy~km~s$^{-1}$)} &
\colhead{(10$^6$ Jy~km~s$^{-1}$ Mpc$^2$)} &
\colhead{(10$^9$ M$_{\sun}$)}
}
\startdata
 NGC 0628 & 6 & 0.57 & \phn 1514 $\pm$ \phn 63\tablenotemark{e}	&  \phn 1.01 $\pm$  0.04\tablenotemark{e}  & 0.63 $\pm$ 0.03\tablenotemark{e} \\ 
 NGC 1068 & 6 & 0.85 & \phn 4102 $\pm$ 217 	& 10.69 $\pm$  0.57  	 & 6.67 $\pm$ 0.35 \\ 
 IC 342   & 6 & 0.28 & \phn 9847\tablenotemark{f} $\pm$ 418 	&  \phn 1.88\tablenotemark{f} $\pm$  0.08  	 & 1.17\tablenotemark{f} $\pm$ 0.05 \\ 
 NGC 2903 & 6 & 0.48 & \phn 3254 $\pm$ 282 	&  \phn 1.62 $\pm$  0.14  	 & 1.01 $\pm$ 0.09 \\ 
 NGC 3351 & 6 & 0.81 & \phn 1513 $\pm$ 221 	&  \phn 1.94 $\pm$  0.28  	 & 1.21 $\pm$ 0.18 \\ 
 NGC 3521 & 6 & 0.55 & \phn 4800 $\pm$ 353 	&  \phn 3.13 $\pm$  0.23  	 & 1.95 $\pm$ 0.14 \\ 
 NGC 3627 & 6 & 0.66 & \phn 4259 $\pm$ 427 	&  \phn 6.59 $\pm$  0.66  	 & 4.12 $\pm$ 0.41 \\ 
 NGC 3938 & 5 & 0.93 & \phn\phn  923 $\pm$ \phn 73 &  \phn 3.35 $\pm$  0.27  	 & 2.09 $\pm$ 0.17 \\ 
 NGC 4258 & 6 & 0.32 & \phn 2686 $\pm$ 255 	&  \phn 2.21 $\pm$  0.21  	 & 1.38 $\pm$ 0.13 \\ 
 NGC 4303 & 6 & 0.92 & \phn 2427 $\pm$ 145 	&  \phn 7.05 $\pm$  0.42  	 & 4.40 $\pm$ 0.26 \\ 
 NGC 4321 & 6 & 0.81 & \phn 2972 $\pm$ 319 	&  \phn 9.68 $\pm$  1.04  	 & 6.04 $\pm$ 0.65 \\ 
 NGC 4414 & 6 & 1.67 & \phn 2453 $\pm$ 282 	& 11.25 $\pm$  1.29  	 & 7.02 $\pm$ 0.81 \\ 
 NGC 4569 & 6 & 0.63 & \phn 1096 $\pm$ 137 	&  \phn 3.89 $\pm$  0.49  	 & 2.43 $\pm$ 0.30 \\ 
 NGC 4736 & 6 & 0.54 & \phn 2641 $\pm$ 155 	&  \phn 0.39 $\pm$  0.05  	 & 0.24 $\pm$ 0.03 \\ 
 NGC 4826 & 6 & 0.60 & \phn 1845 $\pm$ 217 	&  \phn 0.61 $\pm$  0.04  	 & 0.38 $\pm$ 0.02 \\ 
 NGC 5005 & 6 & 1.03 & \phn 1278 $\pm$ 484 	&  \phn 7.29 $\pm$  2.76  	 & 4.55 $\pm$ 1.72 \\ 
 NGC 5033 & 5 & 0.47 & \phn 2469 $\pm$ 258 	& 10.85 $\pm$  1.13  	 & 6.77 $\pm$ 0.71 \\ 
 NGC 5055 & 6 & 0.48 & \phn 3812 $\pm$ 276 	&  \phn 2.48 $\pm$  0.18  	 & 1.55 $\pm$ 0.11 \\ 
 NGC 5194 & 12\tablenotemark{g} & 1.07 & 10097 $\pm$ 267 &  \phn 7.52 $\pm$  0.20  	 & 4.70 $\pm$ 0.12 \\ 
 NGC 5248 & 6 & 0.97 & \phn 1829 $\pm$ 195 	& 11.84 $\pm$  1.26  	 & 7.39 $\pm$ 0.79 \\ 
 NGC 5247 & 5 & 0.89 & \phn 1102 $\pm$ \phn 91 	&  \phn 6.82 $\pm$  0.56  	 & 4.26 $\pm$ 0.35 \\ 
 NGC 5457 & 7 & 0.24 & \phn 3479 $\pm$ 265 	&  \phn 2.39 $\pm$  0.18  	 & 1.49 $\pm$ 0.11 \\ 
 NGC 6946 & 6 & 0.52 & \phn 9273 $\pm$ 241 	&  \phn 3.52 $\pm$  0.09  	 & 2.20 $\pm$ 0.06 \\ 
 NGC 7331 & 6 & 0.57 & \phn 2762 $\pm$ 367 	&  \phn 7.91 $\pm$  1.05  	 & 4.94 $\pm$ 0.66 \\ 
\enddata
\tablenotetext{a}{Angular length of square region over which flux 
density is measured.}
\tablenotetext{b}{Global flux densities measured over square region of 
length $\Delta\alpha$. 
See Figure \ref{boxflux} for flux densities as a function of 
varying $\Delta\alpha$.}
\tablenotetext{c}{CO luminosities, or 4$\pi d^2$ $S_{tot}$, where $d$ is the
distance to the galaxy, listed in Table 1.}
\tablenotetext{d}{CO mass enclosed within region measured. 
$M(H_2)$ = 7845 $S_{tot}$ $d^2$, where $d$ is 
in Mpc, the CO/H$_2$ conversion factor is assumed to be
 2 $\times$ 10$^{20}$ cm$^{-2}$ (K km~s$^{-1}$)$^{-1}$, and
no correction for heavy elements is made.}
\tablenotetext{e}{Formal uncertainties are listed; the systematic error
is 1 $\sigma$ = 15\%.}
\tablenotetext{f}{Size of emitting region is much larger than 6\arcmin\
for IC 342; this flux density should not be interpreted as a global
flux measurement.}
\tablenotetext{g}{Region for NGC 5194 is a rectangle with lengths 
$\Delta\alpha,\Delta\delta=8\arcmin,12\arcmin$.}
\end{deluxetable}

\clearpage

\begin{deluxetable}{lccc}  
\tablecolumns{4}  
\tabletypesize{\footnotesize}
\tablewidth{0pc}  
\tablecaption{Peak and Central Molecular Surface Densities}
\tablehead{  
\colhead{Source} &
\colhead{$\Sigma_{peak}$\tablenotemark{a}} &
\colhead{$\Sigma_{cent}$\tablenotemark{b}} &
\colhead{$\Sigma_{cent}$/$\Sigma_{peak}$}  \\
\colhead{} &
\colhead{(M$_{\sun}$ pc$^{-2}$)} &
\colhead{(M$_{\sun}$ pc$^{-2}$)} &
\colhead{} 
}
\startdata
NGC 0628 & \phn 104 & \phn\phn  20 & 0.20 \\ 
NGC 0925 & \phn\phn  31 & \phn\phn\phn   0 & 0.00 \\ 
NGC 1068 & \phn 947 & \phn 466 & 0.49 \\ 
IC 342 & 1267 & 1151 & 0.91 \\ 
NGC 2403 & \phn\phn  30 & \phn\phn  25 & 0.81 \\ 
NGC 2841 & \phn\phn  36 & \phn\phn\phn   0 & 0.00 \\ 
NGC 2903 & \phn 573 & \phn 573 & 1.00 \\ 
NGC 2976 & \phn\phn  15 & \phn\phn  15 & 1.00 \\ 
NGC 3031 & \phn\phn  24 & \phn\phn\phn   0 & 0.00 \\ 
NGC 3184 & \phn 153 & \phn 153 & 1.00 \\ 
NGC 3344 & \phn\phn  24 & \phn\phn\phn   0 & 0.00 \\ 
NGC 3351 & \phn 426 & \phn 406 & 0.95 \\ 
NGC 3368 & \phn 815 & \phn 645 & 0.79 \\ 
NGC 3521 & \phn 103 & \phn\phn  82 & 0.80 \\ 
NGC 3627 & \phn 539 & \phn 539 & 1.00 \\ 
NGC 3726 & \phn\phn  94 & \phn\phn  94 & 1.00 \\ 
NGC 3938 & \phn\phn  72 & \phn\phn  60 & 0.82 \\ 
NGC 3953 & \phn\phn  23 & \phn\phn\phn   0 & 0.00 \\ 
NGC 3992 & \phn\phn  18 & \phn\phn\phn   5 & 0.32 \\ 
NGC 4051 & \phn 315 & \phn 315 & 1.00 \\ 
NGC 4258 & \phn 473 & \phn 202 & 0.43 \\ 
NGC 4303 & \phn 432 & \phn 432 & 1.00 \\ 
NGC 4321 & \phn 442 & \phn 442 & 1.00 \\ 
NGC 4414 & \phn 130 & \phn\phn  59 & 0.46 \\ 
NGC 4450 & \phn\phn 15 & \phn\phn 10 & 0.71 \\ 
NGC 4490 & \phn\phn 18 & \phn\phn\phn 0 & 0.00 \\ 
NGC 4535 & \phn 429 & \phn 429 & 1.00 \\ 
NGC 4548 & \phn\phn  18 & \phn\phn  18 & 1.00 \\ 
NGC 4559 & \phn\phn  16 & \phn\phn  13 & 0.80 \\ 
NGC 4569 & \phn 510 & \phn 510 & 1.00 \\ 
NGC 4579 & \phn\phn  99 & \phn\phn  89 & 0.90 \\ 
NGC 4699 & \phn\phn 15 & \phn\phn\phn   5 & 0.34 \\ 
NGC 4725 & \phn\phn  69 & \phn\phn  21 & 0.32 \\ 
NGC 4736 & \phn 327 & \phn 327 & 1.00 \\ 
NGC 4826 & \phn 672 & \phn 672 & 1.00 \\ 
NGC 5005 & \phn 724 & \phn 724 & 1.00 \\ 
NGC 5033 & \phn 188 & \phn 188 & 1.00 \\ 
NGC 5055 & \phn 335 & \phn 328 & 0.98 \\ 
NGC 5194 & \phn 787 & \phn 216 & 0.28 \\ 
NGC 5248 & \phn 344 & \phn 323 & 0.94 \\ 
NGC 5247 & \phn 108 & \phn 108 & 1.00 \\ 
NGC 5457 & \phn 348 & \phn 348 & 1.00 \\ 
NGC 6946 & 1854 & 1854 & 1.00 \\ 
NGC 7331 & \phn 164 & \phn\phn  31 & 0.19 \\ 
\enddata
\tablenotetext{a}{Peak face-on molecular surface brightness in the galaxy.}
\tablenotetext{b}{Central face-on molecular surface brightness, measured at the
highest position in the central 6\arcsec.  Central positions taken
from \citet{Sheth02}.}
\end{deluxetable}

\clearpage

\clearpage

\end{document}